%% file: main.tex
\documentclass[11pt]{article}

\usepackage[preprint]{acl}

\usepackage{times}
\usepackage{latexsym}

\usepackage[T1]{fontenc}
\usepackage[utf8]{inputenc}

\usepackage{microtype}

\usepackage{inconsolata}

\usepackage{tabularx}
\usepackage{multirow}
\usepackage{graphicx}
\usepackage{caption}
\usepackage{subcaption}
\usepackage{dblfloatfix}
\usepackage{xcolor}
\usepackage{booktabs}
\usepackage{wrapfig}
\usepackage{enumitem}
\usepackage{longtable}
\usepackage{ragged2e}
\usepackage{array}
\usepackage{amsmath}
\usepackage{pgfplots}
\usepackage{tikz}
\usepackage{siunitx}
\usepackage{color,soul}
\usepackage{cleveref}
\usepackage{makecell}
\usepackage[most]{tcolorbox}
\usepackage{fontawesome5}
\usepackage{pifont}
\usepackage{listings}
\usepackage{tikz}
\usetikzlibrary{arrows.meta,positioning,shadows,shapes.geometric}

\lstdefinestyle{promptstyle}{
    basicstyle=\ttfamily\footnotesize,
    breaklines=true,
    breakatwhitespace=true,
    columns=fullflexible,
    keepspaces=true,
    frame=single,
    rulecolor=\color{gray!40},
    backgroundcolor=\color{gray!5},
    xleftmargin=0.5em,
    xrightmargin=0.5em,
    aboveskip=0.75em,
    belowskip=0.75em
}

\setlist{noitemsep, leftmargin=*, topsep=0pt, partopsep=0pt}
\crefformat{section}{\S#2#1#3} % see manual of cleveref, section 8.2.1
\crefformat{subsection}{\S#2#1#3}
\crefformat{subsubsection}{\S#2#1#3}

\newcommand{\SYSTEM}{BIRDS}

\definecolor{softblue}{rgb}{0.25, 0.45, 0.85}

\newcommand*\circled[1]{\tikz[baseline=(char.base)]{\small{\textbf{
			\node[shape=circle,fill,inner sep=0.75pt] (char) {\textcolor{white}{#1}};}}}}

\newcommand{\cutesquare}[1]{%
    \begin{tikzpicture}[baseline=(char.base)]
        \node[shape=rectangle, rounded corners=2pt, draw=black, fill=black, text=white, inner sep=2pt, font=\small\bfseries] (char) {#1};
    \end{tikzpicture}%
}

\title{\textcolor{cyan}{\faDove} BIRDS: Characterizing and Understanding Biodiversity Impact of Large Language Model Serving}

\author{Tianyao Shi \\
  Purdue University \\
  \texttt{shi676@purdue.edu} \\\And
  Yi Ding \\
  Purdue University \\
  \texttt{yiding@purdue.edu} \\}

\begin{document}
\maketitle

\input{0-abstract}

\input{1-intro}

\input{2-background}
\input{3-method}

\input{4-result}
\input{5-conclusion}

% \clearpage

\input{6-limitation-ethics}

% \begin{acks}
% \end{acks}

% \bibliographystyle{acl}
\bibliography{ref}

\input{appendix}

\end{document}

%% file: 0-abstract.tex
\begin{abstract}

Large language model (LLM) serving creates environmental impacts beyond carbon and water, including ecosystem damage through biodiversity-related pathways. We present \SYSTEM{}, a framework for Biodiversity Impact of Request-Driven LLM Serving. \SYSTEM{} defines request-level functional units, quantifies operational and embodied biodiversity impact, and introduces Quality-Normalized Biodiversity Impact (QNBI) to jointly analyze ecological impact and response quality. Across diverse workloads, models, GPUs, and regions, \SYSTEM{} reveals that biodiversity impact accumulates at scale and exposes quality-aware serving tradeoffs. The code is available at \url{https://github.com/TianyaoShi/BIRDS}.
	
\end{abstract}

%% file: 1-intro.tex
\section{Introduction} \label{sec:intro}

Large language models (LLMs) are rapidly becoming critical computing infrastructure, but their increasing scale and use have also raised significant concerns about environmental impact~\cite{ding2024sustainable}. Recent work has begun to characterize the carbon emissions~\citep{strubell2019energy,li2024sprout,wu2025unveiling} and water consumption~\citep{li2025making,ren2024reconciling} of LLMs. However, environmental impact extends beyond carbon or water alone. Electricity generation and semiconductor manufacturing release air pollutants and chemical waste that contribute to acidification, eutrophication, and ecotoxicity~\cite{blum2024chip,ewaste_impacts}. The ecological stakes are substantial: around 1 million animal and plant species are threatened with extinction and industrial facilities discharge 300--400 million tons of heavy metals, solvents, toxic sludge, and other wastes into waters each year~\citep{ipbes2019global}. Such stressors can damage ecosystems by degrading soil quality, disrupting habitats, and reducing species survival over time~\citep{falk2025more,shi2025servers}.

To provide a more comprehensive understanding of the environmental impact of LLM serving, this paper focuses on characterizing biodiversity impact (BI). BI measures ecosystem damage induced by human activities through multiple environmental pathways~\citep{cardinale2012biodiversity}. Importantly, optimizing LLM serving for carbon or water alone does not necessarily minimize BI. A region with low carbon intensity may still induce high ecological damage through air-pollution-related pathways~\citep{shi2025servers}. As a result, biodiversity-aware serving decisions can differ from carbon-aware or water-aware optimization. Although the BI of a single request is small, modern LLM systems process trillions of tokens per day~\cite{pichai2026io,openrouter2026rankings}, causing ecological impacts to accumulate rapidly at operational scale.

While recent work has modeled BI for computing~\cite{falk2025more,shi2025servers}, existing approaches cannot directly apply to LLM serving for two reasons. \circled{1} They lack a request-driven modeling framework tailored to LLM serving workloads. LLM serving fundamentally operates on requests with serving conditions and latency constraints, yet no serving-oriented functional unit (definition in~\Cref{subsec:lca}) has been defined to attribute BI at the request level. \circled{2} Existing approaches lack mechanisms to reason about the tradeoffs between BI and LLM output quality. Smaller models may appear environmentally preferable while producing lower-quality outputs that require retries or corrections.

\input{figures/bi-intuition}

To address these limitations, we present \SYSTEM{}, a modeling and characterization framework for \textbf{B}iodiversity \textbf{I}mpact of \textbf{R}equest-\textbf{D}riven LLM \textbf{S}erving. The \underline{key insight} behind \SYSTEM{} is that BI accounting for LLM serving should move beyond coarse infrastructure-level estimation into request-driven analysis that explicitly accounts for LLM output quality. In particular, environmentally efficient LLM serving should consider not only the ecosystem damage from generating responses, but also the quality of the responses towards accomplishing the target task. \SYSTEM{} consists of three main steps. First, \SYSTEM{} defines a request-driven functional unit (FU) that models LLM serving under explicit workloads, serving configurations, and latency constraints. Second, \SYSTEM{} quantifies FU-level BI by jointly modeling operational and embodied lifecycle impacts through midpoint-to-endpoint BI accounting. Third, \SYSTEM{} enables quality-aware biodiversity analysis through a new metric, Quality-Normalized Biodiversity Impact (QNBI), which jointly evaluates BI and response quality across models and serving configurations.

We evaluate \SYSTEM{} across diverse LLM serving workloads, model families, and GPU platforms. Our results show that per-request BI is numerically small, but can accumulate rapidly under trillion-token-scale inference traffic; operational impact dominates total BI of LLM serving, making serving efficiency a primary lever for mitigation; BI favors different regions than carbon or water, reshaping sustainability-aware serving decisions; and quality-aware analysis shows that lower per-request BI does not always mean lower overall biodiversity impact when task quality is considered. In particular, intermediate-scale dense models and sparse midxture-of-experts (MoE) models often achieve better QNBI than very small or very large models, though extremely large MoE models can lose this advantage due to multi-GPU overhead. Long-output workloads, reasoning modes and older GPUs can substantially increase BI when their quality or throughput gains do not offset additional serving overhead.

We make the following contributions:
\begin{itemize}
    \item Demonstrate biodiversity impact as a complementary ecological metric beyond carbon and water for LLM serving.
    \item Present \SYSTEM{}, a quality-aware, request-driven framework for characterizing BI of LLM serving.
    \item Define QNBI to jointly compare BI and response quality across models and serving configurations.
    \item Characterize BI across diverse workloads, model families, GPUs, and deployment regions, revealing decision-relevant serving tradeoffs.
\end{itemize}

%% file: figures/bi-intuition.tex
\begin{figure*}[!t]
\centering
% Removed \resizebox to ensure font sizes exactly match your document text
\resizebox{2\columnwidth}{!}{
\begin{tikzpicture}[
    % Set horizontal and vertical spacing explicitly
    node distance=0.5cm and 0.7cm,
    % Modern arrowhead style with sharp lines
    arrow/.style={-{Stealth[scale=1.2]}, ultra thick, draw=black!60},
    % Base style for all blocks to ensure structural consistency
    base/.style={
        rectangle, 
        rounded corners=4pt, 
        draw=black!70, 
        thick,
        align=center, 
        % Fixed width automatically distributes 5 boxes perfectly across \textwidth
        text width=3.2cm, 
        minimum height=1.5cm, 
        font=\small,
        % Adds a professional subtle drop shadow (requires \usetikzlibrary{shadows})
        drop shadow={opacity=0.15, shadow xshift=2pt, shadow yshift=-2pt}
    },
    % A logical color progression mapping infrastructure to ecology
    step1/.style={base, fill=blue!12},
    step2/.style={base, fill=blue!5},
    step3/.style={base, fill=orange!8},
    step4/.style={base, fill=orange!4},
    step5/.style={base, fill=green!12}
]

  % Nodes placed horizontally using relative positioning
  \node[step1] (serving) {
    \textbf{LLM Serving}\\
    \vspace{0.15cm}
    \small Requests, Tokens, GPUs
  };

  \node[step2, right=of serving] (activity) {
    \textbf{Physical Activities}\\
    \vspace{0.15cm}
    \small Energy Use,\\ Hardware Manufacturing
  };

  \node[step3, right=of activity] (stressors) {
    \textbf{Environmental Stressors}\\
    \vspace{0.15cm}
    \small Air Pollution, Toxic Releases
  };

  \node[step4, right=of stressors] (damage) {
    \textbf{Ecosystem Damage}\\
    \vspace{0.15cm}
    \small Acidification, Eutrophycation, Ecotoxicity
  };

  \node[step5, right=of damage] (bi) {
    \textbf{Biodiversity Impact}\\
    \vspace{0.15cm}
    \small Species Loss Over Time
  };

  % Connecting Arrows
  \draw[arrow] (serving) -- (activity);
  \draw[arrow] (activity) -- (stressors);
  \draw[arrow] (stressors) -- (damage);
  \draw[arrow] (damage) -- (bi);

\end{tikzpicture}
}
\caption{Conceptual overview of how LLM serving activities contribute to biodiversity impact. }
\vspace{-0.1in}
\label{fig:bi_intuition}
\end{figure*}
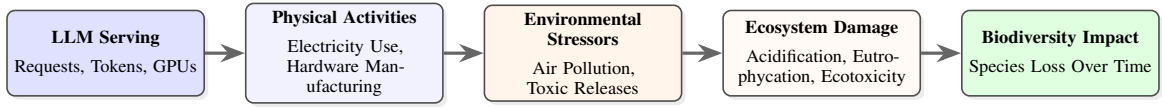

% Serving electricity consumption and hardware manufacturing induce environmental stressors, which further translate into ecosystem damage in biodiversity accounting.

%% file: 2-background.tex
\section{Background and Related Work}\label{sec:background}

% This section first introduces the background on life cycle assessment (LCA) needed to understand BI modeling, followed by related work on environmental impact analysis of LLM serving.

\subsection{Life Cycle Assessment}\label{subsec:lca}

Life cycle assessment (LCA) is a widely used methodology for quantifying the environmental impact induced by products and systems throughout their lifecycle~\cite{finkbeiner2006new}. LCA evaluates environmental stressors through \textbf{midpoint indicators}, such as acidification, eutrophication, and ecotoxicity, which can be further translated into \textbf{endpoint} ecosystem damage metrics. This endpoint damage is expressed in \textbf{species$\cdot$yr}, which estimates ecosystem degradation in terms of species loss integrated over time (100-year horizon by default~\cite{huijbregts2016recipe}). 
% BI is characterized within this LCA framework, since biodiversity degradation is often induced indirectly through multiple environmental pathways rather than a single pollutant or resource type.
As illustrated in~\Cref{fig:bi_intuition}, LLM serving contributes to biodiversity impact through multiple environmental pathways. 

LCA studies commonly distinguish between operational impact and embodied impact~\cite{menzies2007life}. \textbf{Operational impact} refers to the impact induced during system operation, such as energy consumption during LLM serving. \textbf{Embodied impact} captures the lifecycle impact associated with hardware manufacturing, transportation, and end-of-life processes. To enable fair comparison across systems and operating conditions, LCA analyses further rely on a \textbf{functional unit}~\cite{klopffer2014life}, which defines the normalized unit of analysis for comparing environmental impacts across systems or operating conditions.

\subsection{Environmental Impact of LLM Serving}

Recent work has begun to characterize the environmental impact of LLMs focusing on carbon emissions~\cite{strubell2019energy,wu2025unveiling,li2024sprout,faiz2024llmcarbon,shi2025disaggregated,tian2026cache,nguyen2024towards} and water consumption~\cite{li2025making,ren2024reconciling,wu2025not}. However, these analyses do not capture broader biodiversity-related ecological impacts.

Limited recent work has modeled BI for computing systems~\cite{falk2025more,shi2025servers}, focusing on device- and infrastructure-level analyses. However, these approaches cannot directly apply to LLM serving for two reasons. \circled{1} They lack request-driven functional units needed to characterize serving workloads under serving constraints. \circled{2} They lack quality-aware metrics needed to analyze tradeoffs between BI and LLM output quality. In contrast, \SYSTEM{} moves BI accounting from coarse infrastructure-level estimation into request-driven, quality-aware LLM serving analysis.

%% file: 3-method.tex
\section{The \SYSTEM{} Framework}\label{sec:framework}

\input{figures/birds-overview}

This section presents \SYSTEM{}, a modeling and characterization framework for \textbf{B}iodiversity \textbf{I}mpact of \textbf{R}equest-\textbf{D}riven LLM \textbf{S}erving. The key insight behind \SYSTEM{} is that BI accounting for LLM serving should move beyond coarse infrastructure-level estimation into request-driven and quality-aware analysis. As illustrated in \Cref{fig:birds-overview}, \SYSTEM{} consists of three steps. First, \SYSTEM{} establishes a request-driven functional unit (FU) for LLM serving workloads under serving operating points and latency constraints. Second, \SYSTEM{} models both operational and embodied impacts associated with each served FU. Third, \SYSTEM{} performs quality-aware biodiversity analysis across LLM serving configurations and model accuracy.

\subsection{Step 1: Establish a Functional Unit}

% Prior BI accounting framework operates at the infrastructure or device level and does not define serving-oriented functional units~\cite{shi2025servers}. However, LLM serving fundamentally operates on requests with various throughput targets and latency constraints. As a result, BI for LLM serving must be modeled at the request level to enable fair comparison across serving operating points.

Because LLM serving is request-based and constrained by throughput and latency targets, \SYSTEM{} models BI at the request level to enable fair comparison across serving operating points.

\textbf{Functional Unit Definition.}
Assume we model the BI of an LLM serving workload type $w$ executed by a serving instance $i$ in deployment region $r$. A workload type denotes a class of LLM tasks (e.g., summarization or code completion) with similar prompt/response length distribution and service level objective (SLO) with respect to latency. A serving instance denotes the complete serving configuration used to host the model, including the GPU set and parallelism strategy.

Following LCA practice, \SYSTEM{} defines the request-driven functional unit (FU) as one completed request of workload type $w$, served by instance $i$ at throughput $\lambda$ while satisfying latency constraints in terms of Time-to-First-Token (TTFT) and Time-per-Output-Token (TPOT). We denote a serving operating point as $\theta=(w,i,\lambda)$. Here is an example of request-driven FU definition:
\begin{tcolorbox}[
  colback=gray!6,
  colframe=gray!35,
  boxrule=0.4pt,
  arc=2pt,
  left=4pt,
  right=4pt,
  top=3pt,
  bottom=3pt
]
One completed summarization request served by an instance at $\lambda=12$ requests/s, while satisfying TTFT of 2s and TPOT of 150 ms.
\end{tcolorbox}

\subsection{Step 2: Quantify FU-Level BI}\label{sec:bi_modeling}

Following LCA practices (\Cref{subsec:lca}), \SYSTEM{} models BI per FU (BI$_\text{fu}$) as the sum of operational BI (OBI) and embodied BI (EBI). Let $\theta=(w,i,\lambda)$ denote an SLO-satisfying serving operating point:
\begin{align}
\text{BI}_{\text{fu}}(\theta,r)=
\text{OBI}_{\text{fu}}(\theta,r)+
\text{EBI}_{\text{fu}}(\theta).
\end{align}
OBI$_\text{fu}$ captures BI from serving energy consumption, while EBI$_\text{fu}$ captures the amortized lifecycle BI of serving hardware.

\SYSTEM{} adopts midpoint-to-endpoint BI accounting~\citep{huijbregts2016recipe}. Midpoint indicators capture ecological stressors induced by activities. Endpoint indicator translates these stressors into ecosystem damage using ecological characterization factors. This formulation enables \SYSTEM{} to model BI through multiple pathways instead of relying on a single ecological indicator:
\begin{align}
\text{BI}=\sum_{c\in\mathcal{C}} M_c \Phi_c(T),
\quad
M_c=\sum_{q\in\mathcal{Q}} X_q I_{q,c}(r),
\end{align}
where $M_c$ denotes the midpoint impact for category $c$, and $\Phi_c(T)$ converts midpoint impacts into endpoint ecosystem damage by integrating on future time horizon $T$; $q$ denotes an activity or flow, $X_q$ is the attributed amount of that activity, and $I_{q,c}(r)$ is its midpoint intensity for category $c$. For example, when the flow $q$ is operational electricity consumption and the midpoint category $c$ is the terrestrial acidification (TA), $X_q$ is the energy and $I_{q,\mathrm{TA}}(r)$ is the regional pollutant emission intensity in kg SO$_2$eq/kWh. 

\textbf{OBI and EBI Instantiation.}
For OBI, \SYSTEM{} measures energy consumption during steady-state serving and attributes it to completed FUs:
\begin{align}
E_{\mathrm{fu}}(\theta)=
\frac{E(\theta)}{N(\theta)}=
\frac{\bar{P}(\theta)}{\lambda},
\end{align}
where $E(\theta)$ is the measured energy, $N(\theta)$ is the number of completed FUs, $\bar{P}(\theta)$ is the average power and $\lambda$ is the number of completed requests that meet the latency constraints. \SYSTEM{} then converts serving energy into OBI using region-specific grid ecological intensity factors $I_{\mathrm{grid},c}(r)$:
\begin{align}
\mathrm{OBI}_{\mathrm{fu}}(\theta,r) =
E_{\mathrm{fu}}(\theta)
\cdot
\sum_{c\in\mathcal{C}}
I_{\mathrm{grid},c}(r)\Phi_c(T).
\end{align}

For EBI, \SYSTEM{} amortizes the lifecycle EBI of the serving instance over the hardware lifetime:
\begin{align}
\mathrm{EBI}_\mathrm{fu}(\theta) \!\!=\!\! \frac{\mathrm{EBI}_\mathrm{sec}(i)}{\lambda}, \quad
\mathrm{EBI}_\mathrm{sec}(i)\!\!=\!\!\frac{\mathrm{EBI}(i)}{\mathrm{LT}}.
\end{align}
Here, $\mathrm{EBI}_\mathrm{sec}(i)$ and $\mathrm{EBI}(i)$ are the per-second and lifetime EBI of the hardware, respectively. $\mathrm{LT}$ is the assumed lifetime. More details of hardware-specific EBI modeling are in~\Cref{app:gpu_ebi}.

\subsection{Step 3: Analyze Quality-Aware Efficiency}

While $\mathrm{BI}_{\mathrm{fu}}$ models the BI of serving one FU, it does not account for the quality of the generated response. In particular, smaller models may appear environmentally preferable per FU but could produce lower-quality outputs that require retries or human correction.

To enable quality-aware analysis, \SYSTEM{} normalizes $\text{BI}_{\text{fu}}$ by output quality to provide a fair basis of evaluation across models. Based on this intuition, we introduce Quality-Normalized Biodiversity Impact (QNBI). Let $Q(\theta)\in[0,1]$ be the average response quality under serving operating point $\theta$, measured using task-specific evaluation metrics such as exact-match accuracy, pass rate, execution success, or LLM-judge score. We have
\begin{equation}
\mathrm{QNBI}(\theta,r)=
\frac{\mathrm{BI}_{\mathrm{fu}}(\theta,r)}
{\max(Q(\theta),\epsilon)},
\end{equation}
where $\epsilon\!>\!0$ is a small constant used to avoid numerical instability when $Q(\theta)$ approaches zero. A lower QNBI indicates lower BI per functional unit of achieved task-specific output quality, under the same workload and evaluation metric. QNBI values produced by different workloads or quality metrics should not be directly compared.

%% file: figures/birds-overview.tex
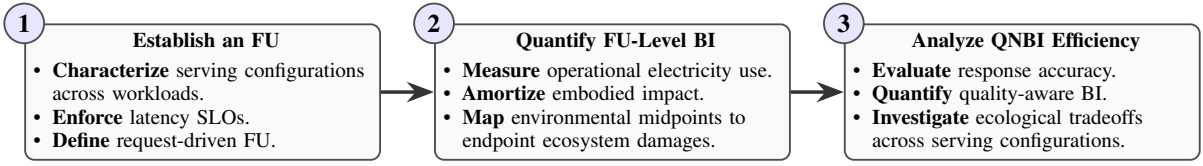
\begin{figure*}[t]
\centering
\resizebox{\textwidth}{!}{
\begin{tikzpicture}[
    node distance=0.8cm,
    % Uniform style for all step boxes
    box/.style={
        rectangle,
        rounded corners=4pt,
        draw=black!70,
        fill=gray!4,
        minimum width=5.2cm,  % Explicitly fixed width
        minimum height=1.8cm, % Explicitly fixed height to guarantee identical sizes
        text width=5.1cm,     % Controlled text bounds
        align=left,           % Left-aligned for proper bullet formatting
        font=\small,
        thick
    },
    % Enlarged step number badge style sitting on the edge
    badge/.style={
        circle,
        draw=black!70,
        fill=blue!10,
        thick,
        font=\bfseries\normalsize, % Increased font size from \small to \normalsize
        minimum size=0.6cm,        % Increased node size from 0.55cm to 0.7cm
        inner sep=0pt,
        anchor=center
    },
    arrow/.style={
        ->,
        line width=1.5pt,
        draw=black!80,
        >={Stealth[scale=1.2]}
    }
]

% ==========================================
% STEP 1: Establish an FU
% ==========================================
\node[box] (step1) {
\vspace{0.4cm}
{\centering\textbf{Establish an FU}\par}
\vspace{-0.3cm} % Pulls the list up immediately below the title
\begin{itemize}
    \setlength{\topsep}{0pt}    % Removes space before the list starts
    \setlength{\itemsep}{0pt}
    \setlength{\parsep}{0pt}
    \item \textbf{Characterize} serving configurations across workloads.
    \item \textbf{Enforce} latency SLOs.
    \item \textbf{Define} request-driven FU.
\end{itemize}
};
\node[badge] at (step1.north west) {1};

% ==========================================
% STEP 2: Quantify FU-Level BI
% ==========================================
\node[box, right=of step1] (step2) {
\vspace{0.4cm}
{\centering\textbf{Quantify FU-Level BI}\par}
\vspace{-0.3cm} % Pulls the list up immediately below the title
\begin{itemize}
    \setlength{\topsep}{0pt}    % Removes space before the list starts
    \setlength{\itemsep}{0pt}
    \setlength{\parsep}{0pt}
    \item \textbf{Measure} operational electricity use.
    \item \textbf{Amortize} embodied impact.
    \item \textbf{Map} environmental midpoints to endpoint ecosystem damages.
\end{itemize}
};
\node[badge] at (step2.north west) {2};

% ==========================================
% STEP 3: Analyze Quality-Aware Efficiency
% ==========================================
\node[box, right=of step2] (step3) {
\vspace{0.4cm}
{\centering\textbf{Analyze QNBI Efficiency}\par}
\vspace{-0.3cm} % Pulls the list up immediately below the title
\begin{itemize}
    \setlength{\topsep}{0pt}    % Removes space before the list starts
    \setlength{\itemsep}{0pt}
    \setlength{\parsep}{0pt}
    \item \textbf{Evaluate} response accuracy.
    \item \textbf{Quantify} quality-aware BI.
    \item \textbf{Investigate} ecological tradeoffs across serving configurations.
\end{itemize}
};
\node[badge] at (step3.north west) {3};

% Connections
\draw[arrow] (step1) -- (step2);
\draw[arrow] (step2) -- (step3);

\end{tikzpicture}
}
\caption{Overview of the three-step modeling procedure of \SYSTEM{}.}
\label{fig:birds-overview}
\end{figure*}

% demonstrating the procedural sequence from establishing the function units to modeling lifecycle footprints and analyzing downstream quality-aware ecological efficiency

%% file: 4-result.tex
\section{Evaluation} \label{sec:evaluation}

We evaluate \SYSTEM{} across diverse LLM serving workloads, models, and serving hardware. We focus on four research questions: \cutesquare{1} What is the overall BI of serving LLM inference requests? \cutesquare{2} Where do BIs originate across operational and embodied lifecycle stages? \cutesquare{3} How does BI relate to or differ from existing environmental metrics such as carbon emission and water consumption? \cutesquare{4} What tradeoffs exist between LLM serving configurations, output quality, and BI? We first describe the evaluation methodology and then present results for each research question.

\subsection{Evaluation Methodology}\label{sec:eval_method}

\textbf{Workloads.}
We evaluate diverse non-agentic serving workloads where requests are self-contained and do not invoke external tools during inference. The evaluated datasets cover everyday text conversation (\citep{sharegpt} and WildChat~\citep{zhao2024wildchat}), knowledge-intensive reasoning (MMLU-Pro~\citep{wang2024mmlu} and SuperGPQA-Hard~\citep{pteam2025supergpqascalingllmevaluation}), IDE-style code completion (CrossCodeEval~\citep{ding2023crosscodeeval} and RepoBench~\citep{liu2023repobench}), and long-context NLP tasks (LongBench~\citep{bai2023longbench}). This workload selection enables us to study how different prompt and response length distributions affect latency, energy consumption, and BI. Detailed dataset descriptions are in \Cref{tab:serving_workload_length_stats} in \Cref{app:exp_specs}.

\textbf{Models.}
To study BI tradeoffs across model capability and scale, we evaluate representative open LLM families, including Llama-2 and 3~\citep{touvron2023llama,grattafiori2024llama}, GPT-OSS~\citep{openai2025gptoss120bgptoss20bmodel}, Qwen3~\citep{qwen3technicalreport}, and Gemma-4~\citep{gemma4}. These models span 0.6B--235B parameters and include both dense and mixture-of-experts (MoE) architectures. This selection enables direct comparison between smaller lower-impact models and larger higher-capability models. Full model variants and compatibility filters are listed in \Cref{tab:evaluated_models} in  \Cref{app:exp_specs}.

\textbf{GPUs.}
We run inference on three datacenter NVIDIA GPUs: L40~\citep{nvidia_l40}, A100~\citep{nvidia_a100}, and H100~\citep{nvidia_h100}. These GPUs differ in memory capacity, bandwidth, and power characteristics, enabling us to study how hardware generation changes throughput, energy consumption, and BI. Detailed GPU specifications are provided in \Cref{tab:gpu_hardware} in \Cref{app:exp_specs}.

\textbf{Metrics.}
We measure throughput, TTFT, TPOT, and GPU power for each workload--model--GPU configuration. We then compute energy consumption and BI$_\text{fu}$ using the modeling in \Cref{sec:bi_modeling} with the latest available annual average grid midpoint intensities of the United States, unless otherwise specified. We compute QNBI where response quality is evaluated using task-specific protocols. Open-ended chat workloads are evaluated by an LLM judge against responses from a reference model, selected as Llama-3.1-8B to provide a balanced comparison basis of model size and capability. Objective benchmarks use their native accuracy metrics. Detailed prompts, scoring rules, and quality scores are provided in \Cref{app:output_quality}.

\textbf{Profiling.}
To obtain realistic measurements, we use vLLM 0.19.1~\citep{kwon2023efficient} as the serving engine and collect GPU power traces through NVML~\citep{nvml}. Following common practice in prior energy characterization studies~\citep{benoit_courty_2024_11171501,chung2026ml}, we focus our operational energy profiling on GPU-board energy, which dominates system energy consumption, and exclude host-side energy from components such as CPUs, DRAM, and storage. For each serving configuration, we sweep incoming request rates and retain only stable operating points that satisfy the SLO constraints in \Cref{app:exp_specs}. Unless otherwise specified, we report results at \textit{maximum sustainable throughput (MST)}, defined as the highest stable SLO-satisfying throughput. MST amortizes serving overheads across the largest number of completed requests, making it the most energy-efficient satisfactory operating point. 
% We also evaluate sub-MST operating points to study how traffic load affects BI$_\text{fu}$ and QNBI. 
Detailed energy profiling methodology is provided in \Cref{app:energy_profiling}.

\subsection{Overall BI Characterization}

\begin{figure}[t]
    \centering
    \includegraphics[width=\linewidth]{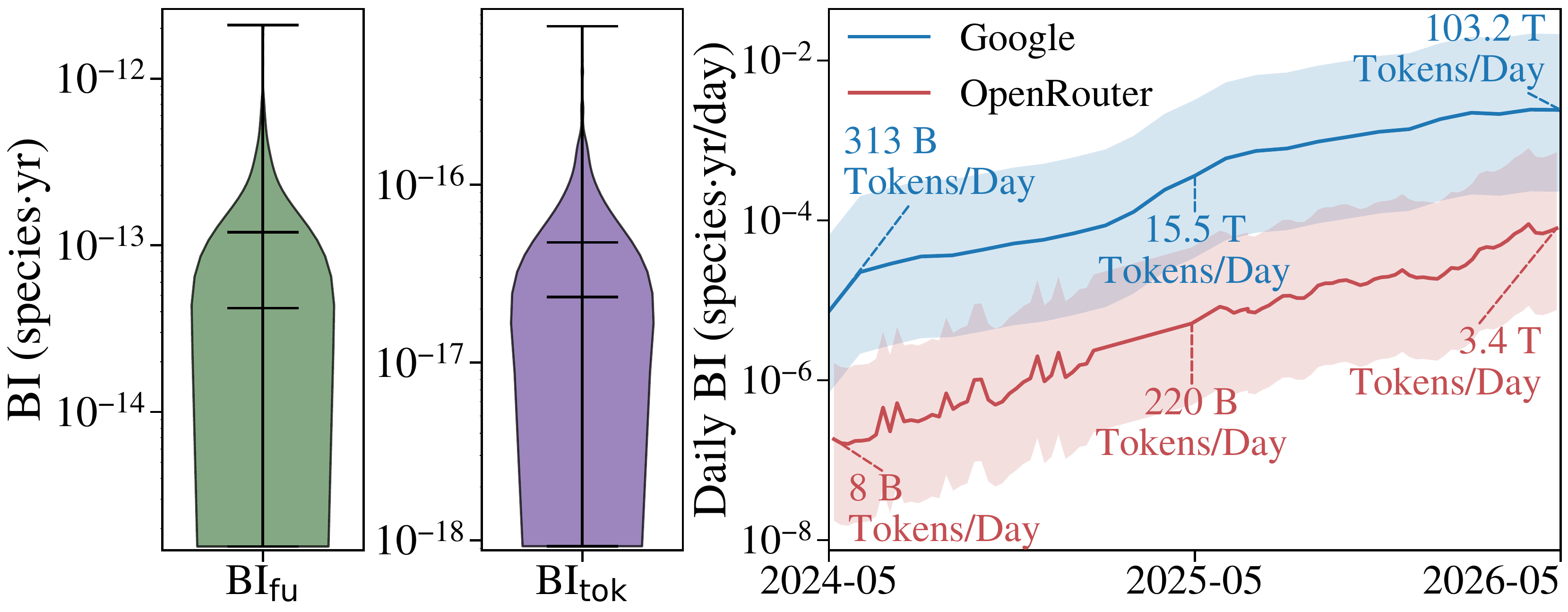}
    \caption{Distribution of FU- and token-level BIs (BI$_\text{fu}$ and BI$_\text{tok}$) across evaluated LLM serving configurations, and daily estimated aggregate BI under publicly reported token traffic from Google and OpenRouter between May 2024 and May 2026. The shaded regions show where 95\% of profiled serving configurations fall.}
    \label{fig:overall_bi}
\end{figure}

We first characterize the overall BI of LLM serving by analyzing both per-request impact and its accumulation under large-scale inference traffic.

% \Cref{fig:overall_bi} summarizes the distribution of request-level and token-level BI across our evaluated serving configurations (left) and further estimates aggregate daily BI under publicly reported LLM token traffic (right).

\textbf{FU- and token-level BIs remain low in magnitude.}
\Cref{fig:overall_bi} (left) shows the distribution of FU- and token-level BIs across profiled serving configurations. At the scale of a single request or token, these values are numerically small, typically corresponding to very small amounts of ecosystem damage. However, modern LLM systems process massive token volumes, causing these impacts to accumulate rapidly at deployment scale.

\textbf{Aggregate BI grows with token traffic.}
To estimate deployment-scale impact magnitude, we multiply the token-level BI distribution by publicly reported token traffic from Google~\citep{pichai2026io} and OpenRouter~\citep{openrouter2026rankings}. Although these traces do not represent the entire LLM ecosystem, they provide observable lower bounds on aggregate inference demand. As shown in \Cref{fig:overall_bi} (right), reported traffic increased from $313$B to $103.2$T tokens/day for the larger trace and from $8$B to $3.4$T tokens/day for the smaller trace between May 2024 and May 2026. By May 2026, the resulting median BI estimates reach $2.41\times10^{-3}$ and $7.95\times10^{-5}$ species$\cdot$yr/day, respectively. Extrapolating $100$T tokens/day over one year yields approximately $0.85$ species$\cdot$yr annually, demonstrating how seemingly negligible per-token impacts can accumulate into non-trivial ecosystem damage at industry scale. For intuition, this annual impact corresponds to approximately \textbf{94,378 single-passenger New York–London round trips} under the same endpoint characterization using the European Reference Life Cycle Database (ELCD) lifecycle inventory data~\cite{greendelta2026elcd}, assuming 11,400 km traveled and 100 kg transported mass per trip. We use this comparison only as an interpretive aid, species$\cdot$yr remains the formal and scientifically meaningful BI metric.

\subsection{BI Source Analysis}
\begin{figure}[t]
    \centering
    \includegraphics[width=\linewidth]{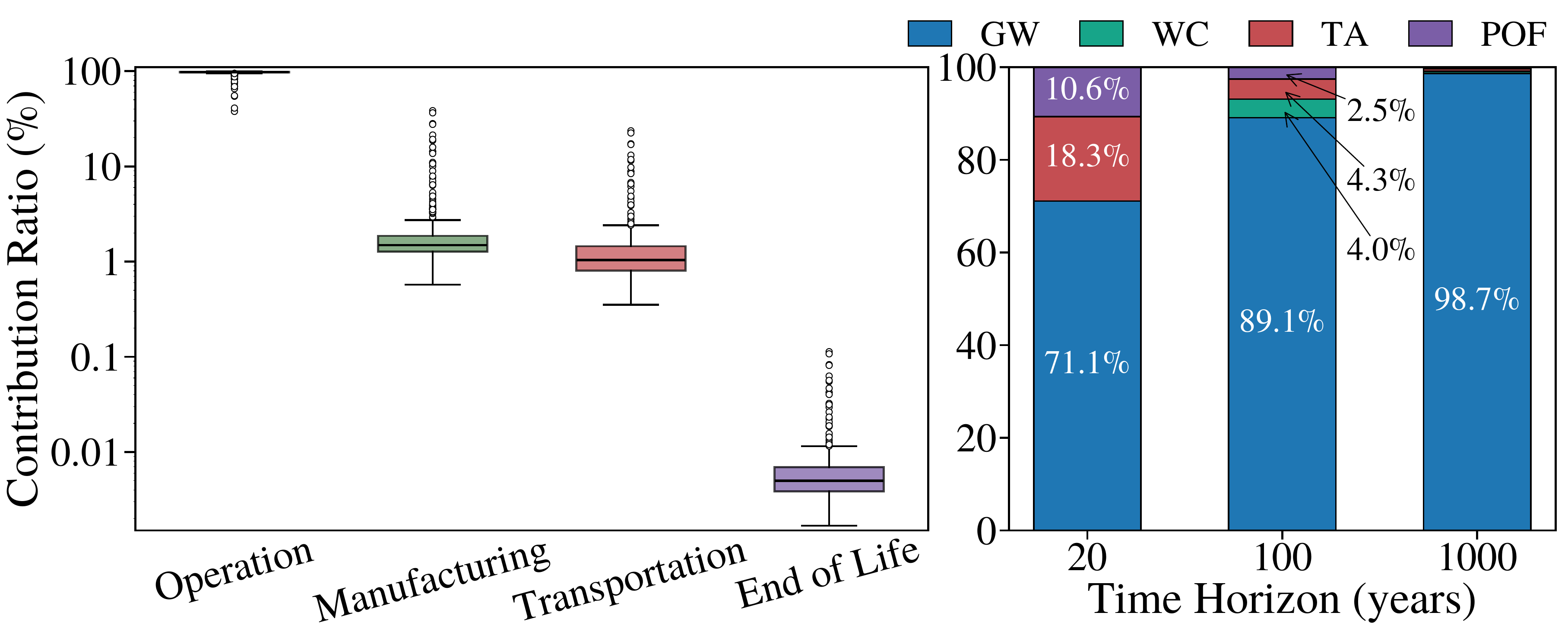}
    \caption{BI$_\text{fu}$ by lifecycle stage integrated over 100 years and by midpoint-to-endpoint contribution ratios under different time horizons. Global warming (GW), water consumption (WC), terrestrial acidification (TA), and photochemical ozone formation (POF) are the top-4 midpoint pathways, with the other midpoints contributing less than 0.01\%.}
    \label{fig:bi_breakdown}
\end{figure}

\begin{table*}[t]
\centering
\footnotesize

\setlength{\tabcolsep}{3.5pt}
\renewcommand{\arraystretch}{1.08}
\vspace{-0.1in}
\begin{tabular}{lccccc}
\toprule
Region &
$I_{\mathrm{grid},\mathrm{GW}}$ &
$I_{\mathrm{grid},\mathrm{WC}}$ &
$I_{\mathrm{grid},\mathrm{TA}}$ &
$I_{\mathrm{grid},\mathrm{POF}}$ &
$\mathrm{OBI}_{1\text{kWh}}$ \\
& kg CO$_2$eq & m$^3$ & kg SO$_2$eq & kg NO$_x$eq & species$\cdot$yr \\
\midrule
Norway & $\mathbf{1.06{\times}10^{-2}}$ & $2.38{\times}10^{-2}$ & $\mathbf{2.08{\times}10^{-5}}$ & $\mathbf{1.60{\times}10^{-5}}$ & $3.57{\times}10^{-10}$ \\
France & $4.02{\times}10^{-2}$ & $6.40{\times}10^{-3}$ & $1.54{\times}10^{-4}$ & $1.53{\times}10^{-4}$ & $\mathbf{2.52{\times}10^{-10}}$ \\
United Kingdom & $1.59{\times}10^{-1}$ & $\mathbf{1.71{\times}10^{-3}}$ & $1.72{\times}10^{-4}$ & $2.14{\times}10^{-4}$ & $5.34{\times}10^{-10}$ \\
California & $1.98{\times}10^{-1}$ & $5.58{\times}10^{-3}$ & $6.52{\times}10^{-5}$ & $1.57{\times}10^{-4}$ & $6.63{\times}10^{-10}$ \\
% Texas & $3.32{\times}10^{-1}$ & $2.37{\times}10^{-3}$ & $2.11{\times}10^{-4}$ & $2.14{\times}10^{-4}$ & $1.03{\times}10^{-9}$ \\
% Japan & $4.29{\times}10^{-1}$ & $4.51{\times}10^{-3}$ & $2.74{\times}10^{-4}$ & $2.98{\times}10^{-4}$ & $1.36{\times}10^{-9}$ \\
\bottomrule
\end{tabular}
\caption{Grid midpoint intensities and operational biodiversity impact (OBI) for consuming 1 kWh of LLM serving energy across regions at the 100-year horizon. The optimal region for each column is highlighted in \textbf{bold}.}
\label{tab:carbon_water_bi_regions}
% \vspace{-0.2in}
\end{table*}

To better understand the ecological drivers of LLM serving, we next analyze where BIs originate across both lifecycle stages and environmental pathways. 
% Specifically, we study whether BI is primarily induced during model serving or hardware production, and which ecological stressors dominate BI under different endpoint time horizons.

\textbf{Operational impact dominates LLM serving BI.}
\Cref{fig:bi_breakdown} (left) decomposes BI$_\text{fu}$ across lifecycle stages under the 100-year horizon. Across profiled serving configurations, the operational stage consistently dominates, contributing more than 95\% of total BI on average. In contrast, manufacturing and transportation contribute at the percent level, while end-of-life impact remains negligible. This result suggests that BI in LLM serving is primarily a use-phase problem driven by electricity consumption rather than hardware disposal. As a result, reducing serving energy consumption and improving serving efficiency are likely more effective for lowering BI than optimizing end-of-life management alone.

\textbf{Different endpoint horizons emphasize different ecological pathways.}
\Cref{fig:bi_breakdown} (right) further decomposes BI into midpoint ecological pathways under different endpoint time horizons. Under average U.S. grid conditions, global warming (GW) is the major contributor, but the relative importance of non-climate pathways changes over time. At the 20-year horizon, GW contributes 71.1\% of total BI, while terrestrial acidification (TA) and photochemical ozone formation (POF) contribute shares of 18.3\% and 10.6\%, respectively. At the 100-year horizon, GW increases to 89.1\%, with water consumption (WC), TA, and POF each contributing only a few percent. At the 1000-year horizon, GW almost fully dominates the endpoint BI, reaching 98.7\%. These results show that BI interpretations depend on the endpoint horizon: longer horizons emphasize climate-related damage, while shorter horizons reveal additional stressors beyond carbon.

\subsection{BI vs. Carbon and Water}

Previous results show that operational impact dominates BI for LLM serving. Because BI captures multiple environmental stressors across power grids and regions, the location with the lowest BI may not be the same as the location favored by carbon or water. This motivates our next question: does BI favor different regions than carbon or water when choosing where to serve LLMs? To answer this question, we measure BI, carbon emissions, and water consumption across multiple regions and compare how each metric ranks the regions.

\textbf{Carbon-, water-, and BI-optimal regions differ.}
\Cref{tab:carbon_water_bi_regions} reports grid midpoint intensities and operational BI for consuming 1 kWh of serving energy across four representative regions: Norway, France, United Kingdom, and California. These regions span diverse electricity mixes and environmental intensity profiles across Europe and North America. The carbon-optimal region is Norway. The water-optimal region is the United Kingdom. However, the BI-optimal region is France. These results show that minimizing carbon or water alone does not necessarily minimize BI, leading BI to favor different regions than carbon or water alone.

\textbf{BI captures tradeoffs across multiple ecological pathways.}
The divergence arises because BI jointly aggregates multiple ecological pathways rather than optimizing a single environmental metric in isolation. Norway achieves the lowest carbon, terrestrial acidification (TA), and photochemical ozone formation (POF) intensities in this region set, but also exhibits the highest water consumption intensity. In contrast, the United Kingdom minimizes water consumption but has substantially higher carbon and air-pollution-related intensities. France is not individually optimal for either carbon or water alone, yet achieves the lowest integrated BI after midpoint-to-endpoint impact conversion. For LLM serving, carbon- or water-only decisions can miss broader ecosystem impacts. BI-aware analysis adds a complementary lens for environmentally responsible serving.

\textbf{Regional distinctions persist across hourly grid variation and endpoint horizons.} To examine whether using annual-average grid intensities hides important temporal variation, we compare regions using hourly 2024 electricity data, sampling 1 day from each quarter (96 hours per region). We obtain hourly generation mix and carbon intensity from ElectricityMaps~\citep{electricitymaps} and use energy-source-specific water, SO$_2$, and NO$_x$ intensities from existing LCA studies~\citep{reig2020guidance,turconi2013life}. Under the default 100-year horizon, the hourly results largely confirm the annual-average findings in \Cref{tab:carbon_water_bi_regions}. Norway is carbon-optimal in 51.0\% of sampled hours, while the United Kingdom remains water-optimal and France remains BI-optimal in every sampled hour. As a result, the BI-optimal region differs from the carbon-optimal region in 51.0\% of hours and from the water-optimal region in 100\% of hours. These results show that hourly changes in the grid do not eliminate the differences between BI-, carbon-, and water-aware regional choices.

We further test whether this result depends on the dates we sample or the endpoint horizon. We sample 4 days per quarter and evaluate all $4^4=256$ quarter-stratified combinations. For each combination, we measure how often carbon and BI select different regions in \Cref{tab:region_sens}. Although the disagreement rate varies across dates and horizons, BI consistently provides a different signal from carbon. At the default 100-year horizon, carbon and BI select different regions for a median of 17.7\% of sampled hours, with disagreement reaching 63.5\% for some date selections. They fully agree in only 24 of the 256 combinations. Together with the persistent disagreement between BI and water above, these results show that BI captures regional ecological tradeoffs that neither carbon nor water alone can represent.

\begin{table}[t]
\centering
\footnotesize
\setlength{\tabcolsep}{3.5pt}
\renewcommand{\arraystretch}{1.05}
\begin{tabular}{lccccc}
\toprule
Horizon & Annual BI & Min & Median & Max & Full  \\
&  optimum &&&&overlap \\
\midrule
20 years & Norway & 16.7\% & 54.2\% & 75.0\% & 0/256 \\
100 years & France & 0.0\% & 17.7\% & 63.5\% & 24/256 \\
1000 years & Norway & 0.0\% & 7.3\% & 33.3\% & 24/256 \\
\bottomrule
\end{tabular}
\caption{Sensitivity of BI–carbon regional disagreement to sampled dates and endpoint horizon. Min, median, and max report the distribution of disagreement rates across the 256 sampled date combinations. ``Full overlap'' counts selections for which BI and carbon choose the same region for all sampled hours.}
\label{tab:region_sens}
\end{table}

\begin{figure*}[t]
    \centering
    \includegraphics[width=\textwidth]{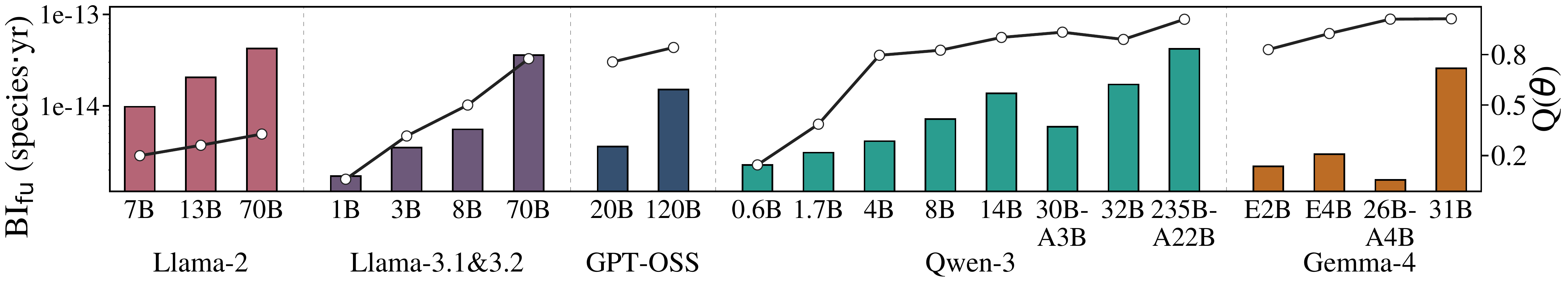}
    \caption{BI$_\text{fu}$ and quality $Q(\theta)$ for everyday chat workload serving. Bars show BI$_\text{fu}$ for the most energy-efficient serving configuration of each model, and the black line shows quality score. Models are grouped by family and ordered by size. For Qwen3 models with Instruct / Thinking variants, we report the Instruct models' results here.}
    \label{fig:raw_bi_qscore}
\end{figure*}

\begin{figure}[t]
    \centering
    \includegraphics[width=\linewidth]{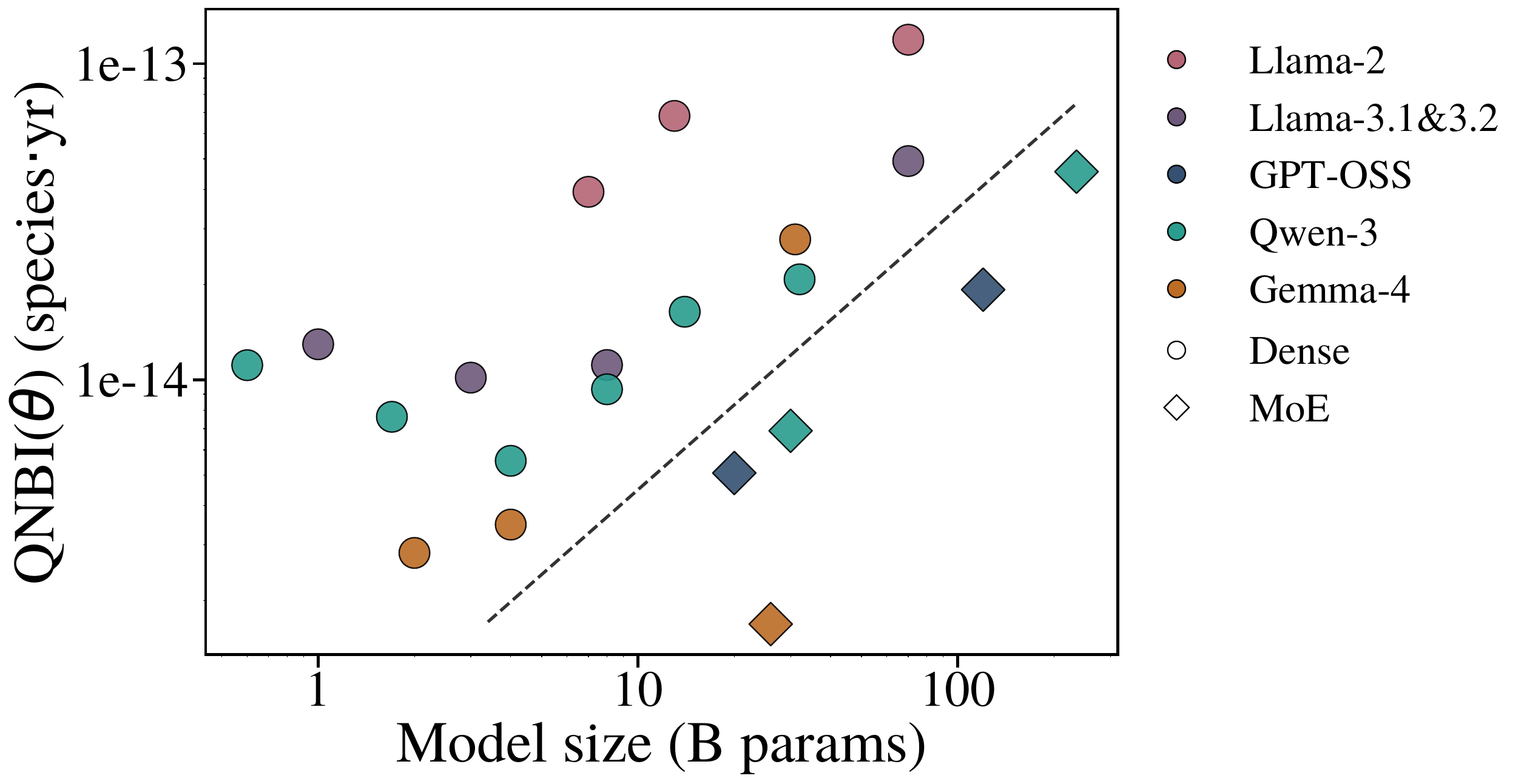}
    \caption{QNBI for daily chat workload serving. Each point corresponds to the most energy-efficient serving configuration of one model. Circles denote dense models and diamonds denote MoE models. The dashed line separates dense models from MoE models.}
    \label{fig:qnbi_scatter_sharegpt}
\end{figure}

\subsection{Quality-Aware BI Analysis}

We next study how BI changes across LLM serving configurations, including model size, model architecture, task type, reasoning mode, traffic load, and GPU generation. Lower BI alone does not necessarily indicate a preferable serving configuration if response quality is too low. Therefore, we jointly analyze BI and QNBI to understand quality-aware ecological efficiency in LLM serving.

\textbf{Model size.}
We first analyze how BI and response quality $Q(\theta)$ change with model size on \citet{sharegpt}, since chatbot interaction is a dominant real-world LLM usage scenario~\cite{chatterji2025people}. \Cref{fig:raw_bi_qscore} shows a tradeoff between response quality and BI as model size increases. Larger models generally achieve higher response quality, but also induce higher BI due to increased computation, memory traffic, and GPU usage. While quality improves rapidly from small to mid-sized models, the gains gradually saturate for recent large models under the LLM-judge  evaluation. In contrast, BI continues to increase with model scale, suggesting that the marginal quality gain of very large models can become relatively small compared to their additional BI for daily chat workloads.

In addition, \Cref{fig:qnbi_scatter_sharegpt} evaluates QNBI across model families. We observe several U-shaped trends within model families, particularly for dense models. Very small models achieve low BI but insufficient response quality, while very large models provide only modest quality gains at higher BI. As a result, the QNBI sweet spot for many dense model families emerges around the 3B--8B scale range. These results suggest that simply choosing the smallest model does not necessarily minimize the BI of LLM serving. Instead, quality-aware BI analysis can identify intermediate operating points that better balance ecological impact and response quality.

\textbf{Model Architecture.} The BI trend in \Cref{fig:raw_bi_qscore} is not strictly monotonic; some MoE models achieve lower BI than smaller dense models because only a subset of parameters is activated, improving the MST greatly. This advantage is more visible in the QNBI results in \Cref{fig:qnbi_scatter_sharegpt}, where MoE models generally achieve lower QNBI than dense models at comparable total model scale. In practice, this suggests that sparse architectures can preserve much of the response-quality benefit of large-scale models while reducing BI during serving. However, the benefit is not unlimited. Extremely large MoE models, such as Qwen3-235B-A22B, still exhibit higher QNBI than some $\sim$30B dense models because serving the full model requires 8 H100 GPUs. In this case, the additional serving energy and hardware overhead are not fully offset by throughput or quality gains. These results suggest that sparse activation can improve ecological efficiency for LLM serving, but very large MoE deployments may still incur substantial ecological impact.

\begin{figure}[t]
    \centering
    \includegraphics[width=\linewidth]{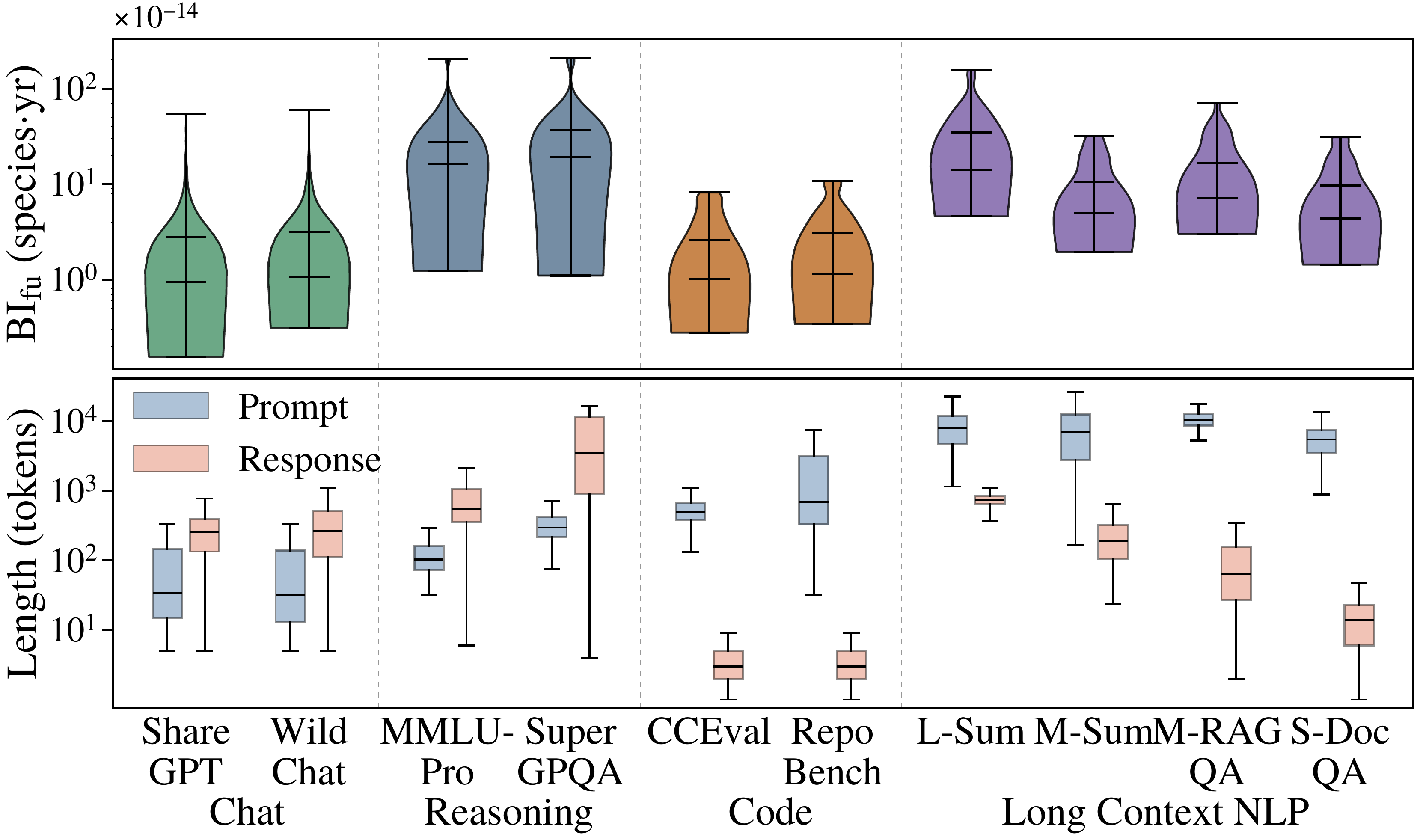}
    \caption{BI$_\text{fu}$ (top) and request length (bottom) distributions across workload types on the H100 GPU. 
    % The upper panel shows BI per FU, and the lower panel shows input and output token lengths. Workloads are grouped into chat, reasoning, code completion, and long context NLP buckets.
    }
    \label{fig:bi_workload_distribution}
\end{figure}

\textbf{Task Type.}
\Cref{fig:bi_workload_distribution} shows that BI varies substantially across LLM workload types, primarily following the amount of computation required per FU. Chat and code-completion workloads generally induce lower BI because their requests are short or output-limited. Code-completion tasks can include large repository contexts, but their generated continuations are typically short, keeping serving time and energy consumption relatively modest.

In contrast, reasoning and long-context workloads induce higher BI because they increase either prompt length, response length, or both. Among them, long response generation is especially costly. SuperGPQA exhibits one of the highest BI distributions because responses can reach up to 10K tokens, substantially extending decoding time and energy consumption. Longer responses also reduce serving throughput since requests retain KV cache for longer time and limit batching efficiency. Long-context workloads exhibit a related but distinct behavior. Large prompts increase prefill computation and KV-cache pressure, while long responses prolong decoding occupancy. As a result, BI depends not only on total token count, but also on how prompt and response lengths interact with serving dynamics. 

\begin{figure}[t]
    \centering
    \includegraphics[width=\linewidth]{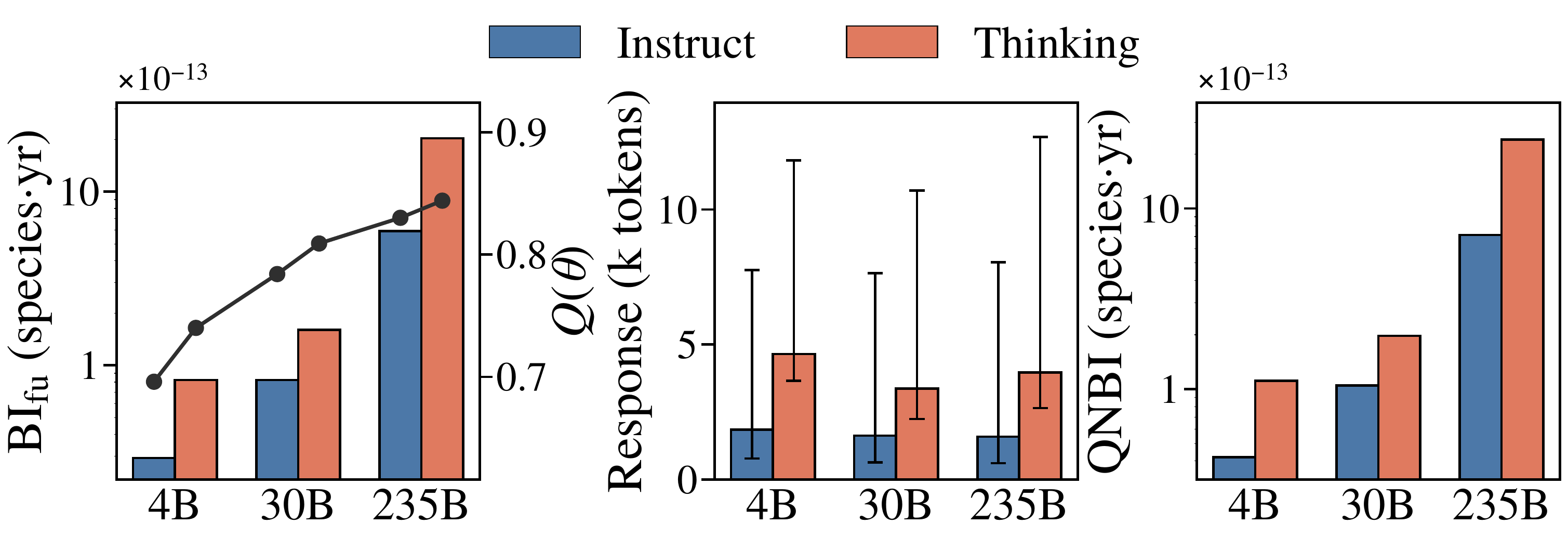}
    \caption{Effect of reasoning mode on MMLU-Pro serving for Qwen3 Instruct and Thinking variants. Left: BI$_\text{fu}$ and quality score \(Q(\theta)\). Middle: response length distribution. Right: QNBI.}
    \label{fig:bi_thinking}
\end{figure}

\textbf{Reasoning Mode.}
We next study the tradeoffs of reasoning-oriented serving using Qwen3 Instruct and Thinking variants on MMLU-Pro. As shown in \Cref{fig:bi_thinking}, Thinking models consistently improve response quality, but also induce substantially higher BI. The main reason is that thinking mode generates much longer responses, roughly doubling or more the response length, which increases decoding time and GPU energy consumption per request. QNBI results further show that the quality gains often do not fully offset the additional ecological impact, especially for larger models such as the 235B variant. These results suggest that reasoning mode should be enabled selectively for workloads where the expected quality improvement justifies the additional serving overhead.

\begin{figure}[t]
    \centering
    \includegraphics[width=\linewidth]{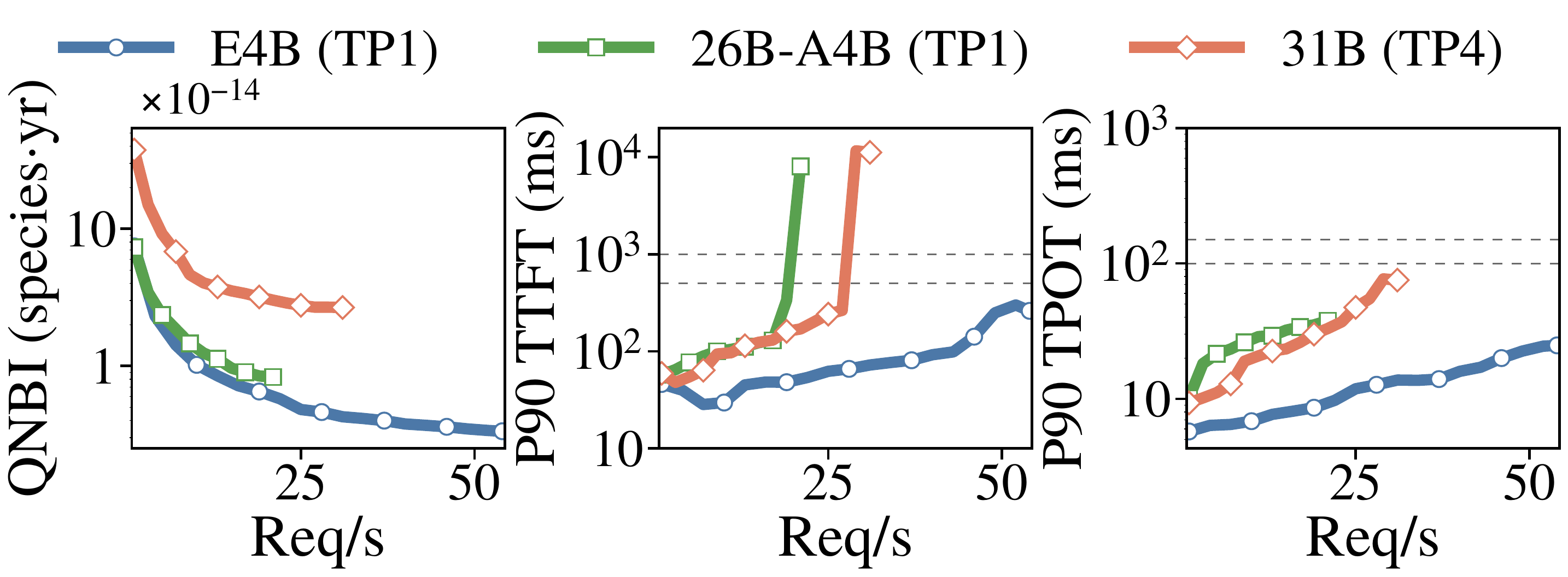}
    \caption{Traffic-load effect on QNBI and latency for ShareGPT serving. Each curve corresponds to one Gemma-4 model and serving configuration. Left: QNBI as incoming request rate increases. Middle and right: p90 TTFT and p90 TPOT; dashed horizontal lines indicate the corresponding SLO thresholds. }
    \label{fig:bi_traffic}
\end{figure}
%\textcolor{red}{Incomplete req rate coverage, relaunching experiments}

\textbf{Traffic Load.} \Cref{fig:bi_traffic} shows that QNBI decreases rapidly as traffic load increases, then gradually flattens near serving saturation. At low load, fixed idle power is amortized over few completed requests, leading to high BI per FU. As load increases, batching and GPU utilization improve, so the same serving instance completes more requests per unit energy and thus QNBI drops. The marginal gain becomes smaller near saturation, where TTFT and TPOT begin to rise and further load increases violates SLO. These results support using MST in our experiments since the most biodiversity-efficient operating point is typically near the highest stable SLO-satisfying throughput.

\begin{figure}[t]
    \centering
    \includegraphics[width=\linewidth]{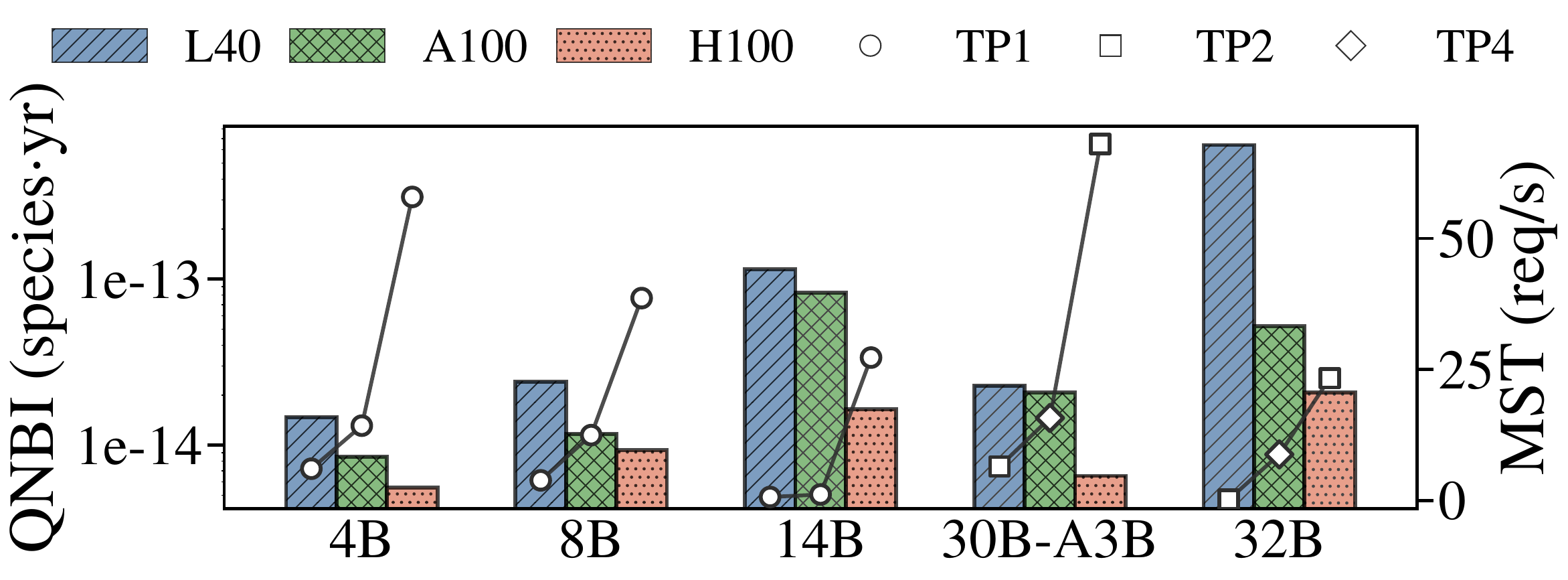}
    \caption{Effect of GPU generation on QNBI and maximum sustainable throughput (MST) for ShareGPT serving. Bar reports QNBI for the most energy-efficient configuration of each Qwen3 model on each GPU. Markers report the corresponding MST, with marker shapes denoting the selected tensor-parallelism (TP) degree.}
    \label{fig:bi_gpu}
\end{figure}

\textbf{GPU Choice.} We next study how GPU generation changes quality-aware BI for LLM serving. \Cref{fig:bi_gpu} compares Qwen3 models from 4B to 32B on ShareGPT across L40, A100, and H100. H100 consistently achieves the lowest QNBI, A100 is usually intermediate, and L40 has the highest QNBI, especially for larger models. This gap grows with model size because larger LLMs require more GPU memory and KV cache capacity. When a model cannot run efficiently on a single device, older GPUs often require more tensor parallelism (TP), which increases energy per request. 
% As a result, the serving configuration becomes less quality-efficient from a biodiversity perspective.

% The MST markers explain this trend. H100 sustains higher request rates because it provides stronger compute, larger memory capacity, higher memory bandwidth, and faster interconnect. This allows serving overheads to be amortized across more completed requests. In contrast, L40 often requires multi-GPU tensor parallelism for larger models, but PCIe communication limits the efficiency gain; the added GPU power is not fully offset by throughput improvements. These results show that GPU choice affects BI not only through power draw, but also through the serving regimes it enables. Older GPUs can remain effective for smaller models that fit on one device and sustain batching, but newer GPUs are more suitable for large or KV-cache-heavy LLM workloads that require high-throughput multi-GPU serving.

The MST markers explain this trend. H100 sustains higher request rates through stronger compute, larger memory capacity, higher bandwidth, and faster interconnect, allowing serving overheads to be amortized across more requests. In contrast, L40 often requires multi-GPU tensor parallelism for larger models, but PCIe communication limits the efficiency gain.  As a result, GPU choice affects BI not only through power draw, but also through the serving regimes it enables: older GPUs remain effective for smaller single-GPU models, while newer GPUs are more suitable for large or KV-cache-heavy workloads.

%% file: 5-conclusion.tex
\section{Conclusion}

We present \SYSTEM{}, a modeling and characterization framework for analyzing the biodiversity impact of request-driven LLM serving. \SYSTEM{} enables quality-aware biodiversity impact accounting and reveals ecological tradeoffs that carbon or water metrics alone cannot capture. Looking forward, \SYSTEM{} provides a foundation for integrating biodiversity impact into the design, deployment, and optimization of environmentally responsible LLM serving systems.

%% file: 6-limitation-ethics.tex
\section*{Limitations}

We discuss the limitations of this work as follows.

\textbf{Non-agentic Scope.} \SYSTEM{} focuses on non-agentic LLM serving workloads where requests are self-contained and do not invoke external tools or retrieval systems. Agentic LLM systems may introduce additional ecological costs through tool usage, multi-step reasoning, external API calls, or repeated interaction loops that are not captured in our current modeling framework.

\textbf{Uncertainty in Biodiversity Accounting.} Our biodiversity accounting relies on life cycle assessment (LCA) characterization factors and regional environmental intensity data. Although these datasets are widely used in environmental assessment, biodiversity modeling remains inherently uncertain due to regional ecological variability, incomplete environmental reporting, and differences across LCA methodologies and endpoint horizons. Therefore, the reported BI values should be interpreted as comparative estimates rather than exact measurements of ecological damage.

% \textbf{Grid Assumptions.} Our operational impact analysis uses average regional grid intensity factors and does not model fine-grained temporal variation in electricity generation. Incorporating spatiotemporal electricity characteristics may further improve biodiversity-aware serving analysis.

\textbf{Quality Evaluation Dependency.} Our quality-aware analysis depends on benchmark-specific evaluation protocols and LLM-judge scoring for open-ended generation tasks. Different evaluation metrics or judge models may affect the QNBI values and ecological tradeoff interpretations.

\textbf{Serving-Focused Lifecycle Coverage.} This work focuses on biodiversity impact during LLM serving and does not model the full lifecycle ecological impact of model training, data collection, or downstream user behavior. Extending biodiversity-aware analysis across the broader lifecycle of foundation models remains important future work.

\section*{Ethical considerations}

This work studies the biodiversity impact of LLM serving systems and aims to support more environmentally responsible AI deployment. Our framework does not involve human subjects, personal data collection, or deployment of LLMs in real-world decision-making settings. All experiments use publicly available models, benchmarks, and serving systems.

The biodiversity impact estimates reported in this work are based on LCA methodologies and should be interpreted as comparative ecological indicators rather than exact measurements of ecosystem damage. Different LCA datasets, regional environmental assumptions, and endpoint horizons may produce different quantitative estimates.

We emphasize that biodiversity impact should complement rather than replace existing environmental metrics such as carbon emissions and water consumption. The goal of this work is not to prescribe a single universal objective, but to encourage broader ecological consideration in future LLM development and deployment decisions.

Quality-aware biodiversity analysis introduces the possibility of optimizing environmental efficiency at the cost of model capability or user experience. We therefore caution against using ecological metrics alone to justify degrading model quality or restricting access to useful AI systems, especially in educational, accessibility, or safety-critical applications.

The authors used AI assistants, including ChatGPT and Codex, during the preparation of this work. They were used for language and LaTeX polishing, public-data crawling and extraction, code generation, debugging, experiment orchestration scripts, and plotting. All AI-assisted text, code, and analysis were reviewed, edited, and validated by the authors. The scientific claims, numerical results, and final interpretations in this paper are based on the authors' modeling choices, profiling experiments, and post-processing scripts; citations were selected and verified by the authors. We did not use AI assistants to process private personal data or reviewer-confidential materials.

\section*{Acknowledgments}
We thank the anonymous reviewers and area chairs for their valuable feedback. We acknowledge Purdue Rosen Center for Advanced Computing for providing A100 GPU hours on Anvil that supported the GPU experiments in this study. 

%% file: appendix.tex
\begin{appendix}

\section{Appendix Overview}\label{app:appendix-overview}

We summarize the appendix as follows:

\begin{itemize}
    \item \Cref{app:lca} details the LCA modeling for biodiversity impact.
    \Cref{app:midpoint} summarizes the ReCiPe2016 midpoint categories used to translate computing-related activities into biodiversity impact. 
    \Cref{app:gpu_ebi} describes how we model embodied biodiversity impact for datacenter GPUs across manufacturing, transportation, and end-of-life stages. 
    \Cref{app:data_source} reports the public data sources used to instantiate operational grid intensities, GPU manufacturing impacts, transportation impacts, and end-of-life impacts.

    \item \Cref{app:exp_specs} lists the evaluation specifications.
    It summarizes the LLM serving workloads, evaluated model families, and GPU platforms used in our experiments. It also defines the workload-specific latency SLOs used to determine valid serving operating points.

    \item \Cref{app:profile_harness} describes the profiling details for LLM serving workloads.
    \Cref{app:output_quality} explains how we measure response quality for open-ended chat workloads and objective benchmarks. 
    \Cref{app:energy_profiling} describes how we collect GPU power traces, compute serving energy, identify stable SLO-satisfying operating points, and determine maximum sustainable throughput.

    \item \Cref{app:additional_results} provides additional results for other workloads.
\end{itemize}

\begin{table*}[t]
  \centering
  \footnotesize
  \setlength{\tabcolsep}{4pt}
  \renewcommand{\arraystretch}{1.15}

  \begin{tabularx}{\textwidth}{@{}
    >{\raggedright\arraybackslash}p{0.23\textwidth}
    >{\raggedright\arraybackslash}p{0.20\textwidth}
    >{\raggedright\arraybackslash}X
    @{}}
    \toprule
    \textbf{Midpoint Impact Category} & \textbf{Unit} & \textbf{Definition} \\
    \midrule

    Global Warming (GW)
      & kg CO$_2$ eq.
      & Species loss from climate-change-driven ecosystem damage caused by greenhouse gas emissions, expressed relative to CO$_2$ equivalents. \\

    Water Consumption (WC)
      & m$^3$
      & Species loss from reduced freshwater availability caused by consumptive water use. \\

    Terrestrial Acidification (TA)
      & kg SO$_2$ eq.
      & Species loss caused by increased soil acidity from emissions such as SO$_2$, NO$_x$, and NH$_3$. \\

    Terrestrial Toxicity (TT)
      & kg 1,4-DCB eq.
      & Species loss from toxic chemical releases to industrial soil, expressed relative to 1,4-dichlorobenzene equivalents. \\

    Photochemical Ozone Formation (POF)
      & kg NO$_x$ eq.
      & Species loss driven by ground-level ozone formation from reactions involving NO$_x$ and volatile organic compounds. \\

    Freshwater Eutrophication (FE)
      & kg P eq.
      & Species loss from phosphorus-driven nutrient enrichment that depletes oxygen in freshwater bodies. \\

    Freshwater Toxicity (FT)
      & kg 1,4-DCB eq.
      & Species loss due to toxic chemical emissions to freshwater, expressed relative to 1,4-dichlorobenzene equivalents. \\

    Marine Eutrophication (ME)
      & kg N eq.
      & Species loss from nitrogen-driven nutrient enrichment in marine environments. \\

    Marine Toxicity (MT)
      & kg 1,4-DCB eq.
      & Species loss from toxic chemical releases to seawater, expressed relative to 1,4-dichlorobenzene equivalents. \\

    \bottomrule
  \end{tabularx}
  \caption{Biodiversity-related midpoint impact categories from the ReCiPe2016~\citep{huijbregts2016recipe} LCA framework used in this paper.}
  \label{tbl:midpoint}
\end{table*}

\section{LCA Modeling for Biodiversity Impact}\label{app:lca}

\begin{table}[t]
  \centering
  \footnotesize
  \setlength{\tabcolsep}{5pt}
  \renewcommand{\arraystretch}{1.12}

  \begin{tabular}{lccc}
    \toprule
    \textbf{Midpoint} & \textbf{$T=20$} & \textbf{$T=100$} & \textbf{$T=1000$} \\
    \midrule
    GW  & $5.32{\times}10^{-10}$ & $2.80{\times}10^{-9}$  & $2.50{\times}10^{-8}$ \\
    WC  & $6.04{\times}10^{-13}$ & $1.35{\times}10^{-8}$  & $1.35{\times}10^{-8}$ \\
    TA  & $2.12{\times}10^{-7}$  & $2.12{\times}10^{-7}$  & $2.12{\times}10^{-7}$ \\
    TT  & $1.14{\times}10^{-11}$ & $1.14{\times}10^{-11}$ & $1.14{\times}10^{-11}$ \\
    POF & $1.29{\times}10^{-7}$  & $1.29{\times}10^{-7}$  & $1.29{\times}10^{-7}$ \\
    FE  & $6.71{\times}10^{-7}$  & $6.71{\times}10^{-7}$  & $6.71{\times}10^{-7}$ \\
    FT  & $6.95{\times}10^{-10}$ & $6.95{\times}10^{-10}$ & $6.95{\times}10^{-10}$ \\
    ME  & $1.70{\times}10^{-9}$  & $1.70{\times}10^{-9}$  & $1.70{\times}10^{-9}$ \\
    MT  & $1.05{\times}10^{-10}$ & $1.05{\times}10^{-10}$ & $1.05{\times}10^{-10}$ \\
    \bottomrule
  \end{tabular}

  \vspace{2pt}
  \begin{minipage}{0.96\linewidth}
  \footnotesize
  For midpoint categories with multiple ecosystem endpoint rows in ReCiPe2016, e.g., GW and WC, $\Phi_c(T)$ sums the corresponding ecosystem-damage factors. 
  \end{minipage}
  \caption{Midpoint-to-endpoint conversion factors $\Phi_c(T)$ from ReCiPe2016~\citep{huijbregts2016recipe}.}
  \label{tab:midpoint_endpoint_factors}
\end{table}

\subsection{Midpoint Categories}\label{app:midpoint}

We detail the environmental impact categories used to translate computing-related physical activities into biodiversity impact. \Cref{tbl:midpoint} reports the ReCiPe2016~\citep{huijbregts2016recipe} midpoint impact categories used in this work, following its ecosystem-quality endpoint characterization factors. It lists each category, the reference unit, and the corresponding ecological interpretation. \Cref{tab:midpoint_endpoint_factors} reports the midpoint to endpoint conversion factor $\Phi_c(T)$ values with respect to different time horizon $T\in\{20,100,1000\}$. 

\subsection{EBI Modeling for Datacenter GPUs}\label{app:gpu_ebi}

Following FABRIC~\cite{shi2025servers}, we model the embodied biodiversity impact of a GPU by decomposing its non-operational life cycle into Manufacturing (Mfg), transportation (Trans), and end-of-life (EoL):
\begin{equation}
    \mathrm{EBI}(d)=
    \sum_{l\in\{\mathrm{Mfg},\mathrm{Trans},\mathrm{EoL}\}}
    \sum_{c\in\mathcal{C}}
    M_{c,l}(d)\Phi_c(T) ,
\end{equation}
where $d$ denotes a GPU, $c$ indexes midpoint categories, $M_{c,l}(d)$ is the midpoint impact of lifecycle stage $l$, and $\Phi_c(r)$ converts midpoint impact into endpoint ecosystem damage.

\paragraph{Manufacturing.}
We model GPU manufacturing as IC production, covering both the accelerator logic dies and attached memory dies. Following ACT~\cite{gupta2022act} and FABRIC~\cite{shi2025servers}, the manufacturing midpoint impact of IC component $x$ is

{\footnotesize
\begin{equation}
    M^{\mathrm{IC}}_{c,\mathrm{Mfg}}(x,r,t)
    =
    \frac{A_x}{Y_x}
    \left(
        \mathrm{EPA}_x \cdot I_{\mathrm{grid},c}(r,t)
        +
        I_{\mathrm{proc},c}(r,t)
    \right),
    \label{eq:gpu_ic_mfg}
\end{equation}
}

where $x$ denotes either a logic or memory IC component, $A_x$ is the active IC area attributed to component $x$, and $Y_x$ is manufacturing yield. $\mathrm{EPA}_x$ is electricity use per unit active silicon area, $I_{\mathrm{grid}c}(r,t)$ is the grid electricity midpoint intensity for category $c$ in region $r$ and year $t$, and $I_{\mathrm{proc},c}(r,t)$ is the process-related midpoint intensity per unit active silicon area (i.e., direct fab pollutant emission in exhaust and wastewater).  We focus on IC components because they dominate semiconductor manufacturing impacts~\cite{falk2025more}, and omit peripheral components like print circuit board (PCB), surface mount device (SMD), and mechanical parts following prior LCA modeling practice~\cite{gupta2022act}.

\paragraph{Transportation.}
Transportation midpoint impact is modeled from shipped mass and transportation mode (e.g., sea, air, truck, etc.):
\begin{equation}
    M_{c,\mathrm{Trans}}(d)=m_d
    \sum_{u\in\mathcal{U}_{\mathrm{trans}}}
    D_u I^{u}_{c,\mathrm{Trans}},
    \label{eq:gpu_trans}
\end{equation}
where $m_d$ is the shipped mass of GPU $d$, $u$ indexes transport modes, $D_u$ is the distance traveled by mode $u$, and $I^{u}_{c,\mathrm{Trans}}$ is the corresponding midpoint intensity per mass-distance product.

\paragraph{End-of-Life.}
End-of-life midpoint impact is modeled using pathway-weighted treatment intensities:
\begin{equation}
    M_{c,\mathrm{EoL}}(d)=m_d
    \sum_{p\in\mathcal{P}_{\mathrm{EoL}}}\rho_p I^{p}_{c,\mathrm{EoL}},
    \quad
    \sum_{p\in\mathcal{P}_{\mathrm{EoL}}}\rho_p=1,
    \label{eq:gpu_eol}
\end{equation}
where $p$ indexes treatment pathways including recycling, incineration, and landfill, $\rho_p$ is the pathway ratio, and $I^{p}_{c,\mathrm{EoL}}$ is the midpoint intensity of pathway $p$. 

\subsection{Data Sources}\label{app:data_source}

We use public and reproducible data sources to parameterize the LCA model. For operational electricity and electricity-related manufacturing impact, the grid midpoint intensity $I_{\mathrm{grid},c}(r,t)$ is derived from authoritative air pollutant emission inventories, including EPA eGRID~\citep{epa2025_egrid2023rev1} and EDGAR~\citep{jrc2024_edgar2024}. The datacenter power usage effectiveness (PUE) and water usage effectiveness (WUE) data are from \citet{microsoft2026efficiency} public release. The operational water consumption data are from existing water studies~\citep{gupta2024dataset}. The midpoint-to-endpoint conversion factors $\Phi_c(r)$ follow the ReCiPe2016  framework~\citep{huijbregts2016recipe}.

For GPU manufacturing, logic-die area $A_x$ is obtained from die-shot analysis or public die-area disclosures~\citep{techpowerup2026h100}, while memory-die area is inferred from memory capacity and bit density~\citep{jones2023modeling}. We set the manufacturing yield $Y_x$ to 0.875, following prior carbon modeling practice~\citep{li2023toward}. The process-related manufacturing midpoint intensity $I^{\mathrm{proc}}_{x,c}(r,t)$, is derived from leading semiconductor vendors' sustainability disclosures, including \citet{TSMC_ESG_Documents} and \citet{SKhynix_Sustainability_Reports}.

For transportation and end-of-life modeling, we use transportation mode, distance assumptions $D_u$, and recycling pathway ratios $\rho_p$ from industrial server LCA reports~\citep{Busa2019_DellR740_LCA}. The transportation midpoint intensities $I^{u}_{c,\mathrm{Trans}}$ and end-of-life pathway intensities $I^{p}_{c,\mathrm{EoL}}$ are taken from the ELCD 3.2 LCA database through OpenLCA~\citep{greendelta2026elcd}. These sources provide the mass-distance and treatment-pathway intensities needed to instantiate \Cref{eq:gpu_trans,eq:gpu_eol}.

\begin{table*}[t]
  \centering
  \footnotesize
  \setlength{\tabcolsep}{2.2pt}
  \renewcommand{\arraystretch}{1.12}

  \begin{tabularx}{\textwidth}{@{}
    >{\raggedright\arraybackslash}p{0.105\textwidth}
    >{\raggedright\arraybackslash}p{0.265\textwidth}
    >{\raggedright\arraybackslash}p{0.235\textwidth}
    *{6}{>{\raggedleft\arraybackslash}p{0.050\textwidth}}
    @{}}
    \toprule
    Task
      & Description
      & Request Dataset
      & \multicolumn{3}{c}{Prompt Length}
      & \multicolumn{3}{c}{Response Length} \\
    \cmidrule(lr){4-6}\cmidrule(l){7-9}
      & & &
      \multicolumn{1}{c}{P50} &
      \multicolumn{1}{c}{P90} &
      \multicolumn{1}{c}{P95} &
      \multicolumn{1}{c}{P50} &
      \multicolumn{1}{c}{P90} &
      \multicolumn{1}{c}{P95} \\
    \midrule

    \multirow[t]{2}{=}{Text Conversation}
      & \multirow[t]{2}{=}{Everyday chatbot conversations sampled from real-world user--LLM interactions.}
      & ShareGPT~\citep{sharegpt}
      & 31 & 701 & 1,377 & 243 & 568 & 703 \\
      & & WildChat~\citep{zhao2024wildchat}
      & 27 & 562 & 1,004 & 233 & 777 & 958 \\
      & & \\

    \midrule
    \multirow[t]{2}{=}{Problem Solving}
      & \multirow[t]{2}{=}{Knowledge-intensive question answering and reasoning workloads.}
      & MMLU-Pro~\citep{wang2024mmlu}
      & 103 & 245 & 317 & \multicolumn{3}{c}{--} \\
      & & SuperGPQA-Hard~\citep{pteam2025supergpqascalingllmevaluation}
      & 296 & 626 & 815 & \multicolumn{3}{c}{--} \\

    \midrule
    \multirow[t]{2}{=}{Code Completion}
      & \multirow[t]{2}{=}{IDE-style code completion requests with repository context or retrieval content embedded in the prompt.}
      & CrossCodeEval~\citep{ding2023crosscodeeval}
      & 490 & 963 & 1,221 & 3 & 8 & 10 \\
      & & RepoBench~\citep{liu2023repobench}
      & 691 & 5,366 & 6,732 & 3 & 7 & 9 \\
      & & \\

    \midrule
    \multirow[t]{4}{=}{Long Context NLP}
      & Long-output, low-interactivity document summarization.
      & LongBench-GovReport~\citep{bai2023longbench}
      & 8,432 & 17,316 & 21,185 & 655 & 876 & 934 \\

      & Medium-output multi-document/news and meeting summarization.
      & LongBench-MultiNews, QMSum, VCSum
      & 2,008 & 11,752 & 16,221 & 160 & 358 & 395 \\

      & Medium-output document-based question answering.
      & LongBench-DuReader
      & 10,396 & 14,258 & 15,196 & 65 & 222 & 263\\

      & Short-output document-based question answering.
      & LongBench-MultiFieldQA, Qasper
      & 5,059 & 8,780 & 10,581 & 14 & 49 & 70 \\

    \bottomrule
  \end{tabularx}
  \caption{Selected LLM serving workloads and their prompt/response length statistics.}
  \label{tab:serving_workload_length_stats}
\end{table*}

\newcolumntype{L}[1]{>{\raggedright\arraybackslash}p{#1}}

\begin{table*}[t]
  \centering
  \footnotesize
  \setlength{\tabcolsep}{2pt}
  \renewcommand{\arraystretch}{1.18}

  \begin{tabularx}{\textwidth}{@{}
    L{0.26\textwidth}
    L{0.12\textwidth}
    L{0.15\textwidth}
    L{0.24\textwidth}
    L{0.16\textwidth}
    @{}}
    \toprule
    \multirow{2}{*}{Family}
      & \multicolumn{4}{c}{Size Class} \\
    \cmidrule(l){2-5}
      & $<2$B
      & 3--8B
      & 13--34B
      & $\geq$70B \\

    \midrule
    Llama 2~\citep{touvron2023llama}
      & --
      & 7B-Chat
      & 13B-Chat
      & 70B-Chat \\

    \midrule
    Llama 3.1/3.2~\citep{grattafiori2024llama}
      & 1B-Instruct
      & 3B-Instruct, 8B-Instruct
      & --
      & 70B-Instruct \\

    \midrule
    GPT-OSS~\citep{openai2025gptoss120bgptoss20bmodel}
      & --
      & --
      & *20B
      & *120B \\

      \midrule

    Qwen3~\citep{qwen3technicalreport}
      & 0.6B, 1.7B
      & 4B-Instruct / Thinking, 8B
      & 14B, *30B-A3B-Instruct / Thinking, 32B
      & *235B-A22B-Instruct / Thinking \\

    \midrule
    Gemma 4~\citep{gemma4}
      & E2B-it
      & E4B-it
      & *26B-A4B-it, 31B-it
      & -- \\

    \bottomrule
  \end{tabularx}

  \vspace{2pt}
  \begin{minipage}{0.98\textwidth}
    \footnotesize
    \emph{Note.} Asterisks mark MoE models; all other listed models are dense models.
    MoE models are bucketed by total parameter count. Qwen3 models with Instruct / Thinking variants are the latest 2507 version. 
  \end{minipage}
  \caption{Evaluated model families grouped by parameter-size bucket.}
  \label{tab:evaluated_models}
\end{table*}

\begin{table*}[t]
  \centering
  \footnotesize
  \setlength{\tabcolsep}{6pt}
  \renewcommand{\arraystretch}{1.15}

  \begin{tabularx}{\textwidth}{@{}
    >{\raggedright\arraybackslash}p{0.25\textwidth}
    >{\centering\arraybackslash}p{0.13\textwidth}
    >{\raggedright\arraybackslash}p{0.16\textwidth}
    >{\raggedright\arraybackslash}p{0.22\textwidth}
    >{\raggedright\arraybackslash}X
    @{}}
    \toprule
    GPU & Architecture & Memory & Memory Bandwidth & TDP \\
    \midrule

     L40~\citep{nvidia_l40}
      & Ada Lovelace
      & 48 GB GDDR6
      & 864 GB/s
      & 300 W \\

    A100 SXM~\citep{nvidia_a100}
      & Ampere
      & 40 GB HBM2
      & 1,555 GB/s
      & 400 W \\

    H100 SXM~\citep{nvidia_h100}
      & Hopper
      & 80 GB HBM3
      & 3.35 TB/s
      & Up to 700 W \\

    \bottomrule
  \end{tabularx}
  \caption{NVIDIA GPU platforms used in our LLM serving experiments.}
  \label{tab:gpu_hardware}
\end{table*}

\section{List of Workloads, Models, and GPUs}\label{app:exp_specs}

\Cref{tab:serving_workload_length_stats} lists the LLM serving workloads and their corresponding request datasets with prompt and response length statistics used in the measurement studies in this work. \Cref{tab:evaluated_models} summarizes the evaluated model families and their parameter-size buckets. \Cref{tab:gpu_hardware} list the specifications of GPU platforms used in the experiments.

Latency SLOs specify the service condition under which a request-level functional unit is defined. A serving operating point is considered valid only when the achieved throughput satisfies the workload-specific p90 TTFT and p90 TPOT constraints. \Cref{tab:slo_static} summarizes the consolidated SLOs by workload family and model-size bucket. Chat workloads use interactive latency targets; code workloads use tighter TPOT constraints for completion responsiveness; reasoning workloads allow longer generation latency; and LongBench uses a prompt-length-scaled TTFT threshold because input lengths vary substantially across tasks.

\begin{table}[t]
\centering
\footnotesize

\setlength{\tabcolsep}{4pt}
\renewcommand{\arraystretch}{1.08}
\begin{tabular}{llcc}
\toprule
Workload & Model size & TTFT & TPOT \\
\midrule
\multirow{3}{*}{Text Conversation}
  & $<3$B & 250 ms & 50 ms \\
  & 3B--4B & 500 ms & 100 ms \\
  & $\geq$7B & 1000 ms & 150 ms \\
\midrule
\multirow{3}{*}{Problem Solving}
  & $<3$B & 1500 ms & 150 ms \\
  & 3B--4B & 1500 ms & 150 ms \\
  & $\geq$8B & 2000 ms & 200 ms \\
\midrule
\multirow{3}{*}{CrossCodeEval}
  & $<3$B & 500 ms & 40 ms \\
  & 3B--4B & 750 ms & 50 ms \\
  & $\geq$8B & 1500 ms & 75 ms \\
\midrule
\multirow{3}{*}{RepoBench}
  & $<3$B & 5000 ms & 50 ms \\
  & 3B--4B & 5000 ms & 50 ms \\
  & $\geq$8B & 5000 ms & 75 ms \\

\midrule
Long-Context NLP
  & all buckets & length-scaled & 150 ms \\
\bottomrule
\end{tabular}

\caption{Consolidated latency SLOs used, reported as p90 TTFT and p90 TPOT.}
\label{tab:slo_static}
\end{table}

For LongBench, a static TTFT threshold would either over-constrain long-prompt tasks or over-relax shorter document QA tasks. We therefore use a length-scaled TTFT threshold. For a request $i$ in LongBench subtask $b$ with prompt length $L_i$ input tokens, the TTFT threshold is

\begin{equation}
    T^{\mathrm{TTFT}}_{i,b}
    =
    1000\cdot
    \min\left(
        C_b,\;
        B_b + \alpha_b \frac{L_i}{1000}
    \right)
    \quad \mathrm{ms},
    \label{eq:longbench_length_scaled_ttft}
\end{equation}

where $B_b$ is the base TTFT budget in seconds, $\alpha_b$ is the additional budget per 1K input tokens, and $C_b$ is the cap in seconds. All LongBench subtasks use a fixed p90 TPOT threshold of 150 ms.

\begin{table}[t]
\centering
\footnotesize

\setlength{\tabcolsep}{4pt}
\renewcommand{\arraystretch}{1.08}
\begin{tabular}{lccc}
\toprule
LongBench subtask & $B_b$ (s) & $\alpha_b$ (s / 1K tok.) & $C_b$ (s) \\
\midrule
Long Summarization & 8.0 & 1.4 & 45.0 \\
Medium Summarization & 6.0 & 1.2 & 40.0 \\
Medium RAG QA & 5.0 & 0.9 & 30.0 \\
Short Document QA & 4.0 & 0.8 & 20.0 \\
\bottomrule
\end{tabular}
\caption{Length-scaled TTFT parameters for LongBench workloads. The active TTFT threshold is computed by \Cref{eq:longbench_length_scaled_ttft}; TPOT is fixed at 150 ms for all LongBench subtasks.}
\label{tab:longbench_slo}
\end{table}

\section{Profiling Harness}\label{app:profile_harness}

\subsection{Output Quality}\label{app:output_quality}

We evaluate output quality using two tracks: a chat-style LLM-as-judge track and a ground-truth benchmark track. For open-ended chat workloads, we compute a pairwise quality score against a chosen reference response:
\begin{equation}
    q_{\mathrm{chat}}
    =
    \frac{N_{\mathrm{win}} + 0.5N_{\mathrm{tie}}}
    {N_{\mathrm{win}} + N_{\mathrm{tie}} + N_{\mathrm{loss}}},
\end{equation}
where a win means the judge prefers the candidate response, a tie receives half credit, and invalid judgments are excluded from the denominator. Listing~\ref{lst:llm_judge_prompt} shows the LLM-judge prompt template we used. The reference model is selected as Llama-3.1-8B and the evaluator model is GPT-5.4. We randomize A/B ordering when constructing judge batches and keep position-wise aggregation statistics to check for judge-position bias. This score is used as $Q(\theta)$ for chat-style workloads in QNBI.

\begin{lstlisting}[
    style=promptstyle,
    caption={Pairwise LLM-judge prompt for everyday text conversation quality evaluation.},
    label={lst:llm_judge_prompt}
]
You are an impartial judge comparing two assistant responses to the same user request.

User request:
{prompt}

Assistant A:
{response_a}

Assistant B:
{response_b}

Judge which response better satisfies the user request.

For objective or technical prompts, prioritize factual correctness, reasoning correctness, and functional correctness.
For subjective or open-ended prompts, consider helpfulness, relevance, factual soundness, clarity, and conciseness.
Do not prefer a response merely because it is longer or more structured in terms of formatting.
If both responses are similarly good or similarly flawed, choose Tie.

Return JSON only:
{
  "winner": "A" | "B" | "Tie",
  "reason": "one concise sentence"
}
\end{lstlisting}

For objective benchmarks, we use each benchmark's native or adapter-selected quality metric. MMLU-Pro and SuperGPQA are scored by final-answer accuracy against the ground-truth label. Their results are collected from each model's official technical report or Huggingface model card when they are available. For Llama and Gemma models missing SuperGPQA score, we run the benchmarks following the original code. LongBench tasks are scored with the original task-to-metric mapping and reported primarily by workload bucket: ROUGE-style scores for summarization and Chinese QA (DuReader) tasks, and F1-style scores for document QA tasks.   For CrossCodeEval, we use edit similarity (ES) as the primary score, with exact match (EM) and identifier F1 as supporting metrics. For RepoBench, we report ES as the primary score and EM as the companion metric.The selected primary score for each workload is used as $Q(\theta)$ (ROUGE-style scores are divided by 100). 

\Cref{tab:quality_scores} list the exact quality score values of studied models from our quality profiling experiments. LongBench summarization scores should be interpreted with caution: its ROUGE-style metrics often reward compact responses that closely match the reference wording, so larger instruction-tuned models may be penalized for producing more verbose or hedged answers even when the summary is substantively acceptable.

\begin{table*}[t]
\centering
\scriptsize

\setlength{\tabcolsep}{2.4pt}
\renewcommand{\arraystretch}{1.05}
\resizebox{\textwidth}{!}{
\begin{tabular}{lrrrrrrrrrrrr}
\toprule
Model & Chat & MMLU-& Super& \multicolumn{3}{c}{CrossCodeEval} & \multicolumn{2}{c}{RepoBench} & \multicolumn{4}{c}{LongBench} \\
\cmidrule(lr){5-7}\cmidrule(lr){8-9}\cmidrule(lr){10-13}
 &  & Pro & GPQA & ES & EM & ID-F1 & ES & EM & L-Sum & M-Sum & RAG-QA & Doc-QA \\
\midrule
Llama-2-7B & 24.9 & 20.3 & 12.8  & 51.8 & 6.8 & 40.7 & 30.3 & 0.1  & -- & -- & -- & -- \\
Llama-2-13B & 30.1 & 25.3 &  14.7 & 52.9 & 7.8 & 42.1 & 34.4  & 0.01  & -- & -- & -- & -- \\
Llama-2-70B & 35.6 & 37.5 &  18.0  & 53.7 & 9.5 & 44.0 &  31.6 & 1.62  & -- & -- & -- & -- \\
\addlinespace[1pt]
\midrule
Llama-3.2-1B & 13.3 & 22.2 &  11.6 & 46.5 & 4.4 & 33.9 & 24.9  &  0.2 & 30.4 & 21.7 & 26.9 & 32.6 \\
Llama-3.2-3B & 34.7 & 11.9 &  18.5 & 49.5 & 6.5 & 38.3 & 38.3  &  3.2 & 36.1 & 22.8 & \textbf{37.1} & 49.7 \\
Llama-3.1-8B & 50.0 & 44.2 &  21.0 & 52.3 & 9.1 & 42.4 &  43.5 & 8.2  & 36.6 & \textbf{23.9} & 34.6 & 53.5 \\
Llama-3.1-70B & 73.0 & 62.8 &  35.4 & 54.9 & 11.6 & 45.9 &  56.0 & 24.8  & \textbf{37.7} & 23.8 & 32.6 & 55.1 \\
\addlinespace[1pt]
\midrule
GPT-OSS-20B & 71.4 & 73.6 & 44.9 & -- & -- & -- & -- & -- & 30.9 & 18.6 & 19.8 & 35.0 \\
GPT-OSS-120B & 78.5 & 80.8 & 51.9 & -- & -- & -- & -- & -- & 28.8 & 17.8 & 22.0 & 35.0 \\
\addlinespace[1pt]
\midrule
Qwen3-0.6B & 20.4 & 24.7 & 15.0 & 46.8 & 4.8 & 34.5 &  36.5 & 0.2  & 28.1 & 21.0 & 33.0 & 30.7 \\
Qwen3-1.7B & 40.5 & 36.8 & 20.9 & 49.2 & 6.6 & 38.2 &  39.6 & 4.6  & 31.3 & 22.3 & 33.2 & 41.0 \\
Qwen3-4B-I & 74.7 & 69.6 & 42.8 & 52.4 & 9.4 & 42.3 &  44.6 &  9.2 & 30.9 & 20.5 & 26.4 & 51.1 \\
Qwen3-4B-T & 62.3 & 74.0 & 47.8 & -- & -- & -- & -- & -- & 33.5 & 21.3 & 24.5 & 53.5 \\
Qwen3-8B & 77.1 & 56.7 & 31.6 & 52.4 & 10.0 & 42.7 &  46.5 & 12.2  & 33.5 & 21.8 & 26.9 & 51.8 \\
Qwen3-14B & 83.5 & 61.0 & 34.3 & 53.1 & 10.6 & 43.8 & 58.1  &  22.3 & 33.6 & 22.1 & 31.0 & 53.4 \\
Qwen3-30B-A3B-I & 86.1 & 78.4 & 53.4 & 54.8 & 12.2 & 46.0 &  51.1 &  11.5 & 31.6 & 19.9 & 24.6 & 53.0 \\
Qwen3-30B-A3B-T & 80.8 & 80.9 & 56.8 & -- & -- & -- & -- & -- & 31.1 & 20.6 & 27.6 & 53.0 \\
Qwen3-32B & 82.5 & 65.5 & 39.8 & 53.4 & 11.4 & 44.2 & 58.0 & 20.5 & 33.3 & 22.0 & 29.5 & 53.2 \\
Qwen3-235B-A22B-I & 92.5 & 83.0 & 62.6 & 57.1 & 15.0 & 48.9 &  \textbf{66.7} & \textbf{34.7}  & 32.5 & 20.1 & 25.6 & 53.6 \\
Qwen3-235B-A22B-T & 87.9 & 84.4 & \textbf{64.9} & -- & -- & -- & -- & -- & 32.4 & 18.8 & 23.5 & 52.6 \\
\addlinespace[1pt]
\midrule
Gemma-4-E2B & 77.5 & 60.0 & 32.4 & 42.4 & 3.1 & 29.5 &  34.6 &  1.85 & 32.7 & 21.7 & 33.5 & 49.4 \\
Gemma-4-E4B & 85.4 & 69.4 &  40.4 & 53.9 & 6.5 & 41.4 &  43.6 &  8.39 & 32.0 & 21.6 & 27.2 & 54.0 \\
Gemma-4-26B-A4B & 92.5 & 82.6 &  56.9 & 54.9 & 9.2 & 45.2 & 39.4 & 5.6 & 32.9 & 20.9 & 26.2 & 54.7 \\
Gemma-4-31B & \textbf{92.7} & \textbf{85.2} & 62.3 & \textbf{65.3} & \textbf{16.5} & \textbf{54.8} & 52.2  & 18.1  & 32.1 & 21.7 & 25.0 & \textbf{56.2} \\
\bottomrule
\end{tabular}
}
\caption{Output-quality scores used for QNBI. All values are percentages. The best scores are in bold for each benchmark.  Chat denotes the LLM-judge win/tie quality score; MMLU-Pro and SuperGPQA denote benchmark accuracy; CrossCodeEval reports edit similarity (ES), exact match (EM), and identifier F1 (ID-F1); RepoBench reports edit similarity and exact match; LongBench reports bucket-level scores for long-output summarization (L-Sum), medium-output summarization (M-Sum), medium-answer RAG QA, and short-answer document QA. Missing entries indicate incompatible model--workload pairs: Llama~2 models are excluded from long-context tasks due to their 4K context-length limit; Reasoning models (GPT-OSS and Qwen3-Thinking variants) are excluded from code-completion tasks.}
\label{tab:quality_scores}
\end{table*}

Quality-generation runs use deterministic single-sample decoding with \verb|temperature|=0.0. SuperGPQA uses a 16K-token completion budget, LongBench uses 4K, and both code-completion benchmarks use 128 tokens. For models with explicit thinking modes, we disable thinking for non-thinking model variants where supported, and remove visible reasoning traces before scoring always-thinking variants when required by the benchmark protocol.

For IDE-style code-completion workloads, retrieval context is materialized offline rather than executed live. RepoBench and CrossCodeEval prompts serialize the already-available repository context, imports, and current-file prefix into a single request payload; no tool call, retrieval step, or agent loop occurs during serving. For compatibility with modern instruction-tuned models, some reruns use a chat-formatted completion prompt that marks the cursor position and asks the model to return only the code continuation. This changes the delivery format but not the benchmark semantics: each request remains a non-agentic single-turn completion task with fixed embedded context.

\subsection{Energy Profiling}\label{app:energy_profiling}

We measure GPU-board energy during serving by sampling NVML power counters at 25 ms intervals and numerically integrating the power trace over the measurement window. For multi-GPU serving instances, we collect traces from all visible GPUs and sum their energy. The energy measurement is then multiplied by the datacenter PUE to account for extra overhead like cooling at the infrasturacture level. Each profiling run includes an idle-baseline period, request-traffic warmup, steady-state measurement, and cooldown. Unless otherwise noted, we use 30 s idle-baseline measurement, 60 s traffic warmup, 180 s profiling duration, 15 s cooldown, and three repeated trials with 30 s between-repeat cooldown.

To reduce boundary artifacts, we truncate the first and last 5 s of GPU-monitor traces before computing energy statistics. Runtime serving metrics, including throughput, TTFT, TPOT, and request completion statistics, are recorded at 1 s intervals. We evaluate stability over sliding 10 s windows and retain only operating points that complete requests without queue divergence and satisfy the workload-specific TTFT and TPOT constraints.

For each workload--model--GPU setting, we sweep incoming request rates to identify stable SLO-satisfying operating points. The maximum sustainable throughput (MST) is the largest achieved throughput that remains stable under the latency constraints. Since fixed serving overheads and idle power are amortized over more completed requests at higher stable throughput, MST typically gives the lowest energy and biodiversity impact per request among satisfactory operating points. Sub-MST runs are retained for the traffic-load analysis.

\section{Additional Results}
\label{app:additional_results}

\begin{figure*}[t]
    \centering
    \includegraphics[width=\textwidth]{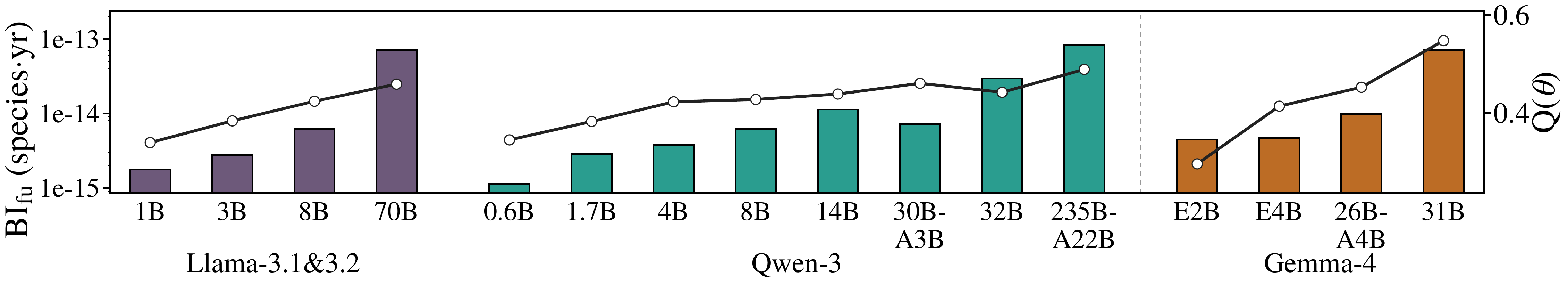}
    \caption{BI$_\text{fu}$ and quality score for CrossCodeEval.}
    \label{fig:app_crosscodeeval_bi}
\end{figure*}

\begin{figure*}[t]
    \centering
    \includegraphics[width=\textwidth]{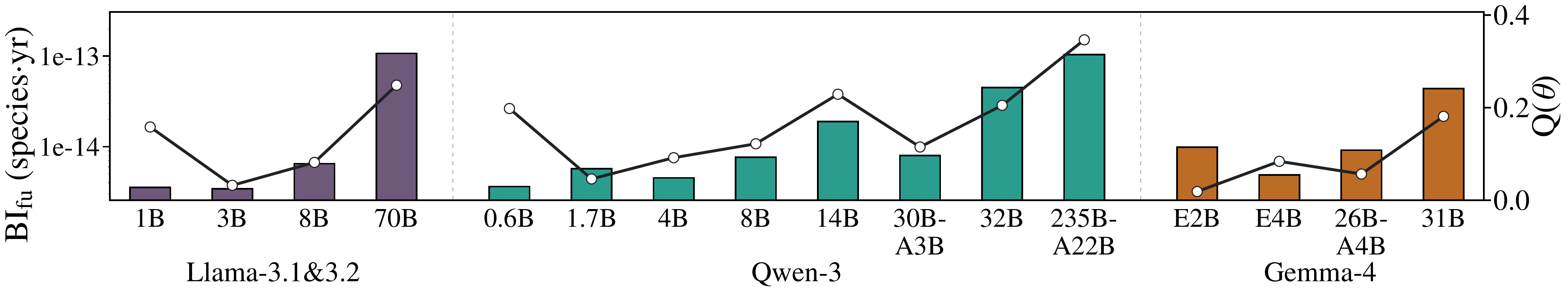}
    \caption{BI$_\text{fu}$ and quality score for RepoBench.}
    \label{fig:app_repobench_bi}
\end{figure*}

\begin{figure*}[t]
    \centering
    \includegraphics[width=\textwidth]{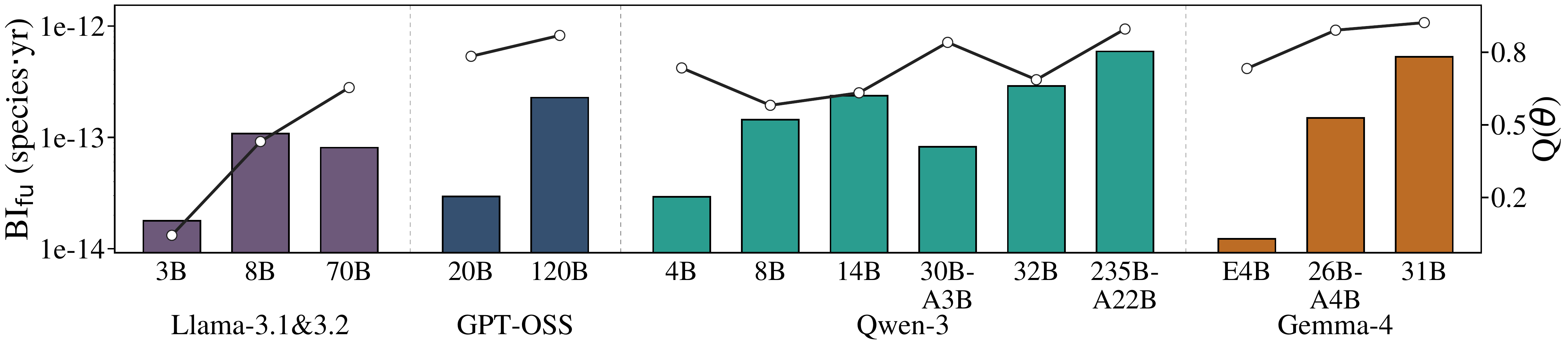}
    \caption{BI$_\text{fu}$ and quality score for MMLU-Pro.}
    \label{fig:app_mmlu_pro_bi}
\end{figure*}

\begin{figure*}[t]
    \centering
    \includegraphics[width=\textwidth]{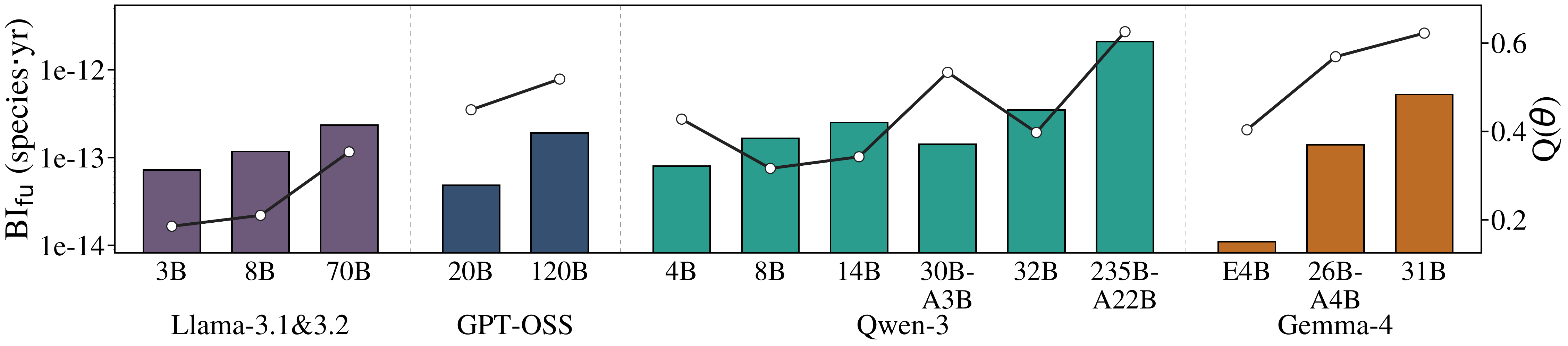}
    \caption{BI$_\text{fu}$ and quality score for SuperGPQA.}
    \label{fig:app_supergpqa_bi}
\end{figure*}

\begin{figure*}[t]
    \centering
    \includegraphics[width=\textwidth]{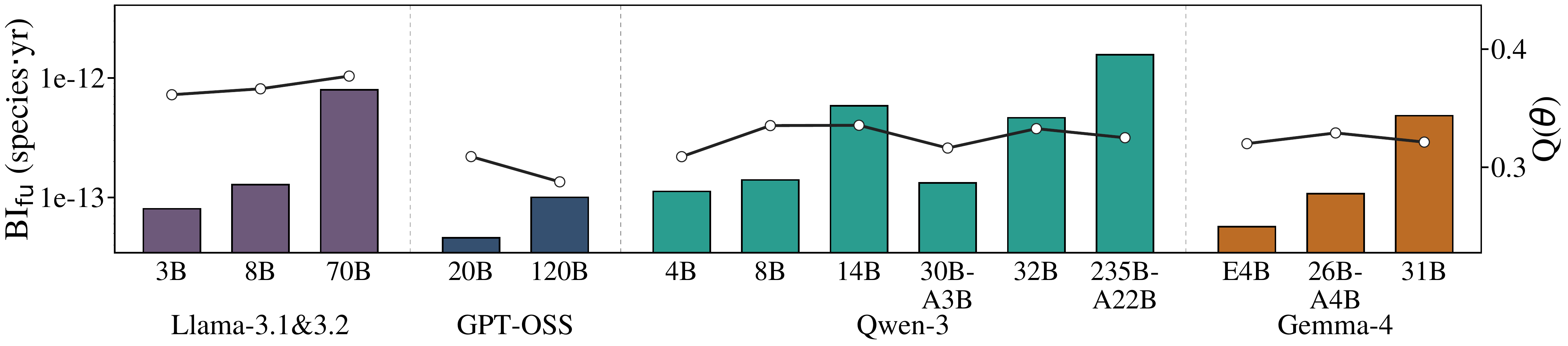}
    \caption{BI$_\text{fu}$ and quality score for LongBench long-output summarization.}
    \label{fig:app_lb_long_sum_bi}
\end{figure*}

\begin{figure*}[t]
    \centering
    \includegraphics[width=\textwidth]{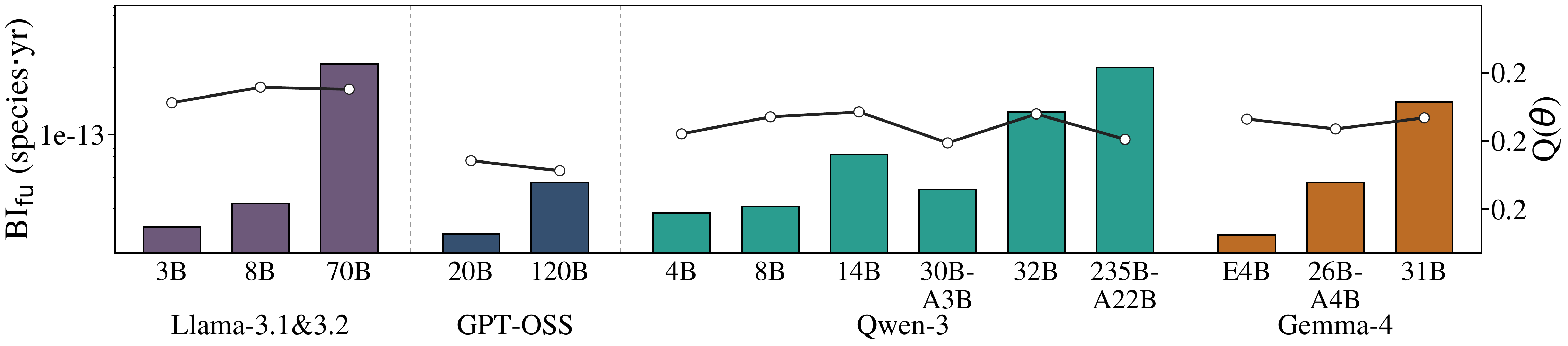}
    \caption{BI$_\text{fu}$ and quality score for LongBench medium-output summarization.}
    \label{fig:app_lb_medium_sum_bi}
\end{figure*}

\begin{figure*}[t]
    \centering
    \includegraphics[width=\textwidth]{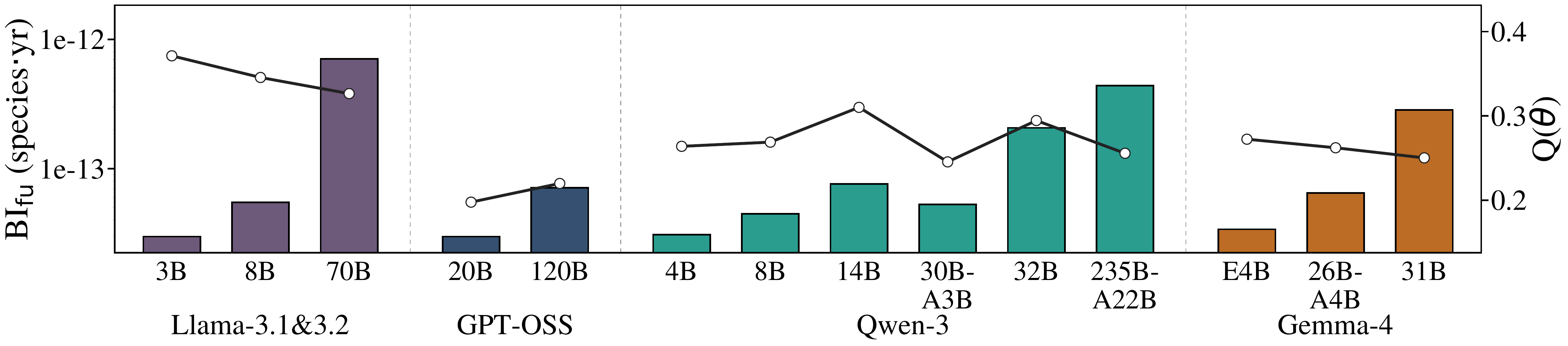}
    \caption{BI$_\text{fu}$ and quality score for LongBench medium-answer RAG QA.}
    \label{fig:app_lb_rag_qa_bi}
\end{figure*}

\begin{figure*}[t]
    \centering
    \includegraphics[width=\textwidth]{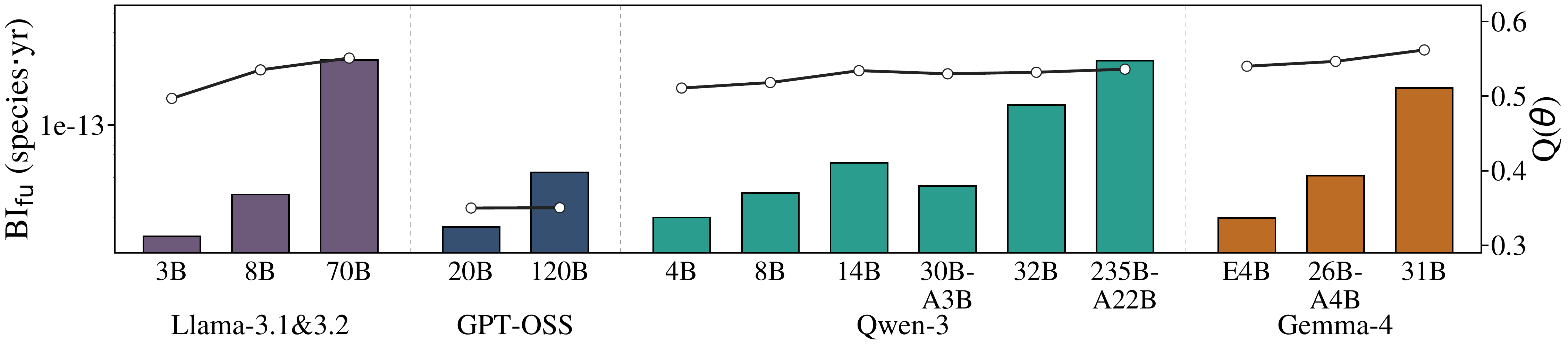}
    \caption{BI$_\text{fu}$ and quality score for LongBench short-answer document QA.}
    \label{fig:app_lb_short_docqa_bi}
\end{figure*}

\begin{figure}[t]
    \centering
    \includegraphics[width=\linewidth]{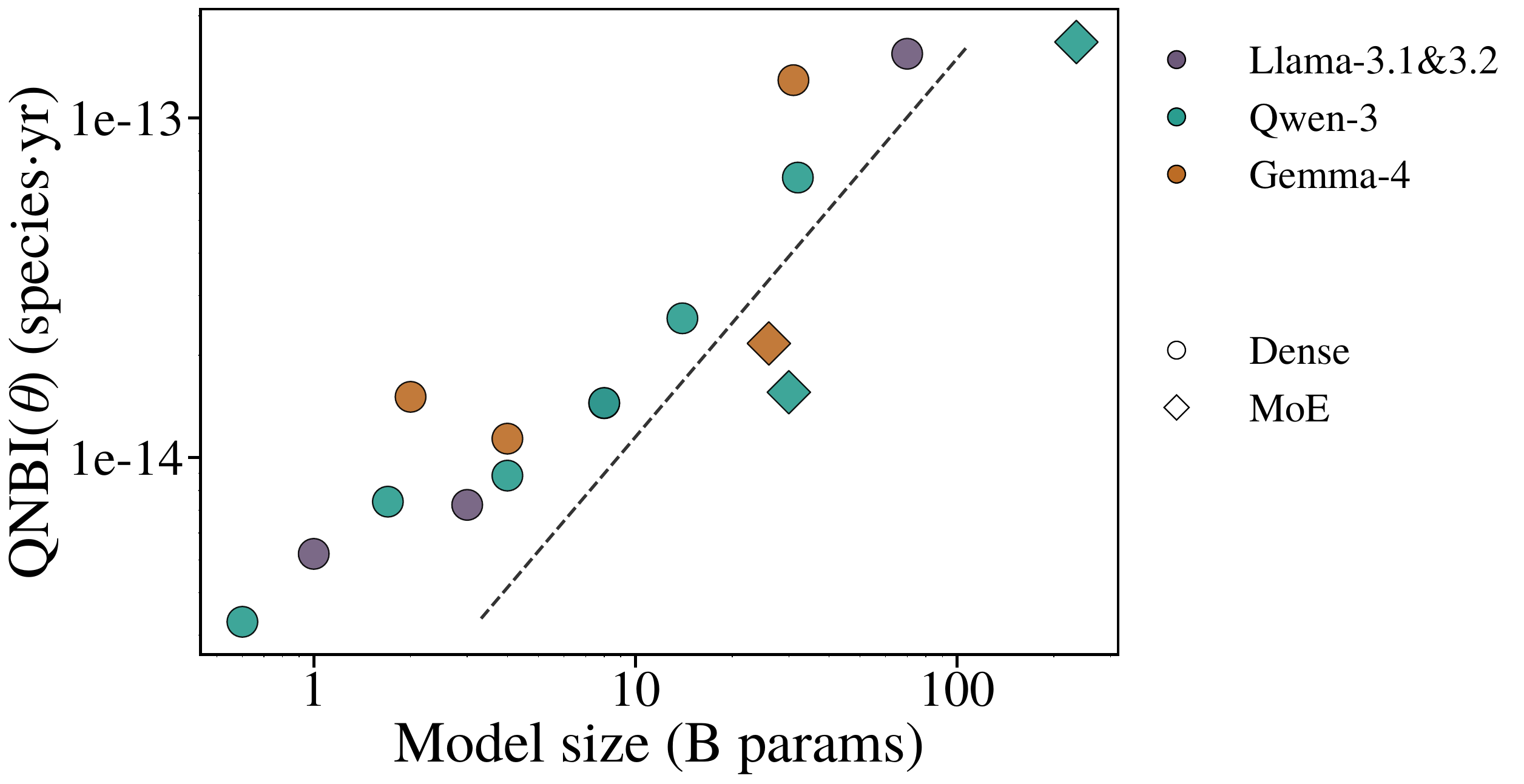}
    \caption{QNBI for CrossCodeEval.}
    \label{fig:app_crosscodeeval_qnbi}
\end{figure}

\begin{figure}[t]
    \centering
    \includegraphics[width=\linewidth]{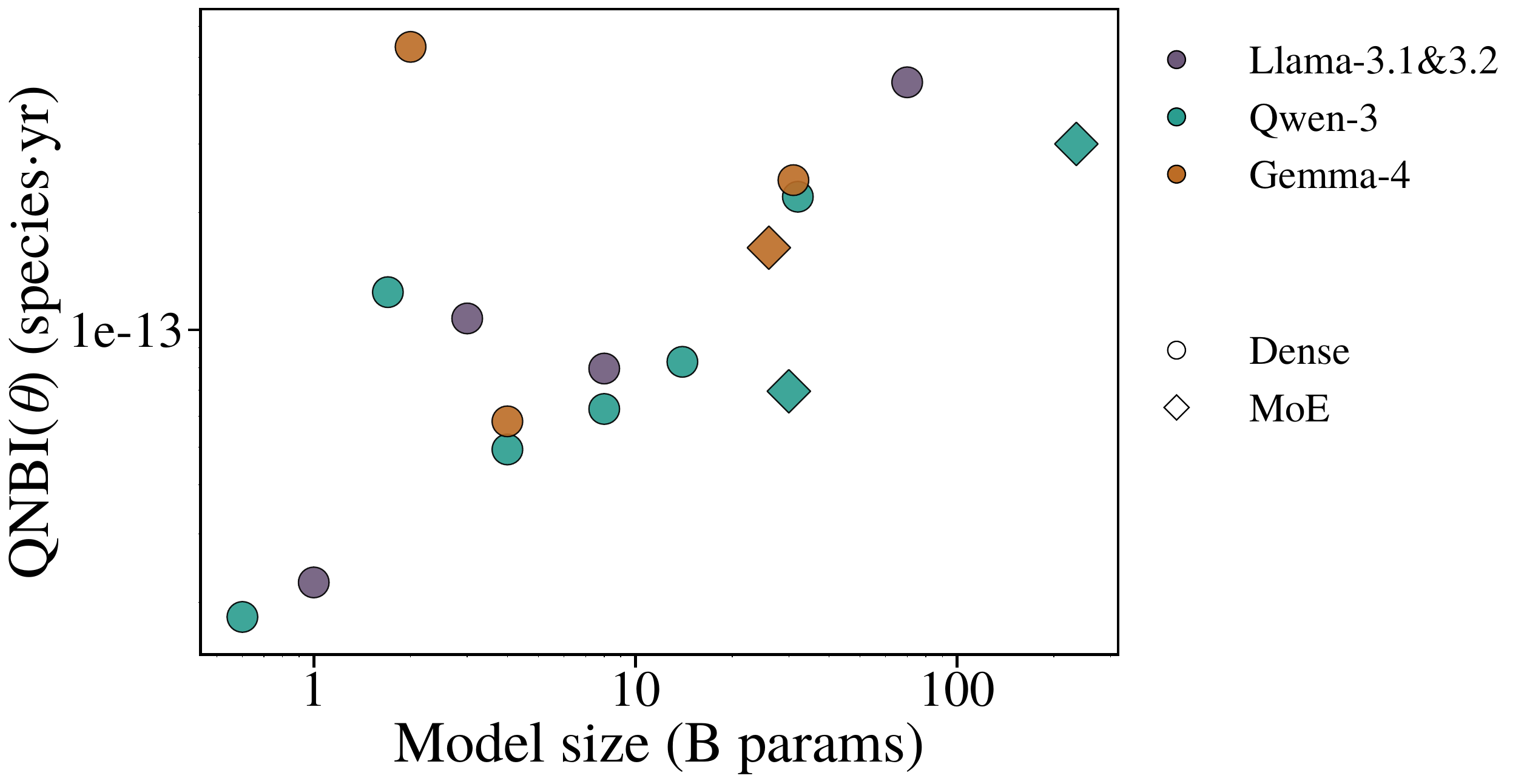}
    \caption{QNBI for RepoBench.}
    \label{fig:app_repobench_qnbi}
\end{figure}

\begin{figure}[t]
    \centering
    \includegraphics[width=\linewidth]{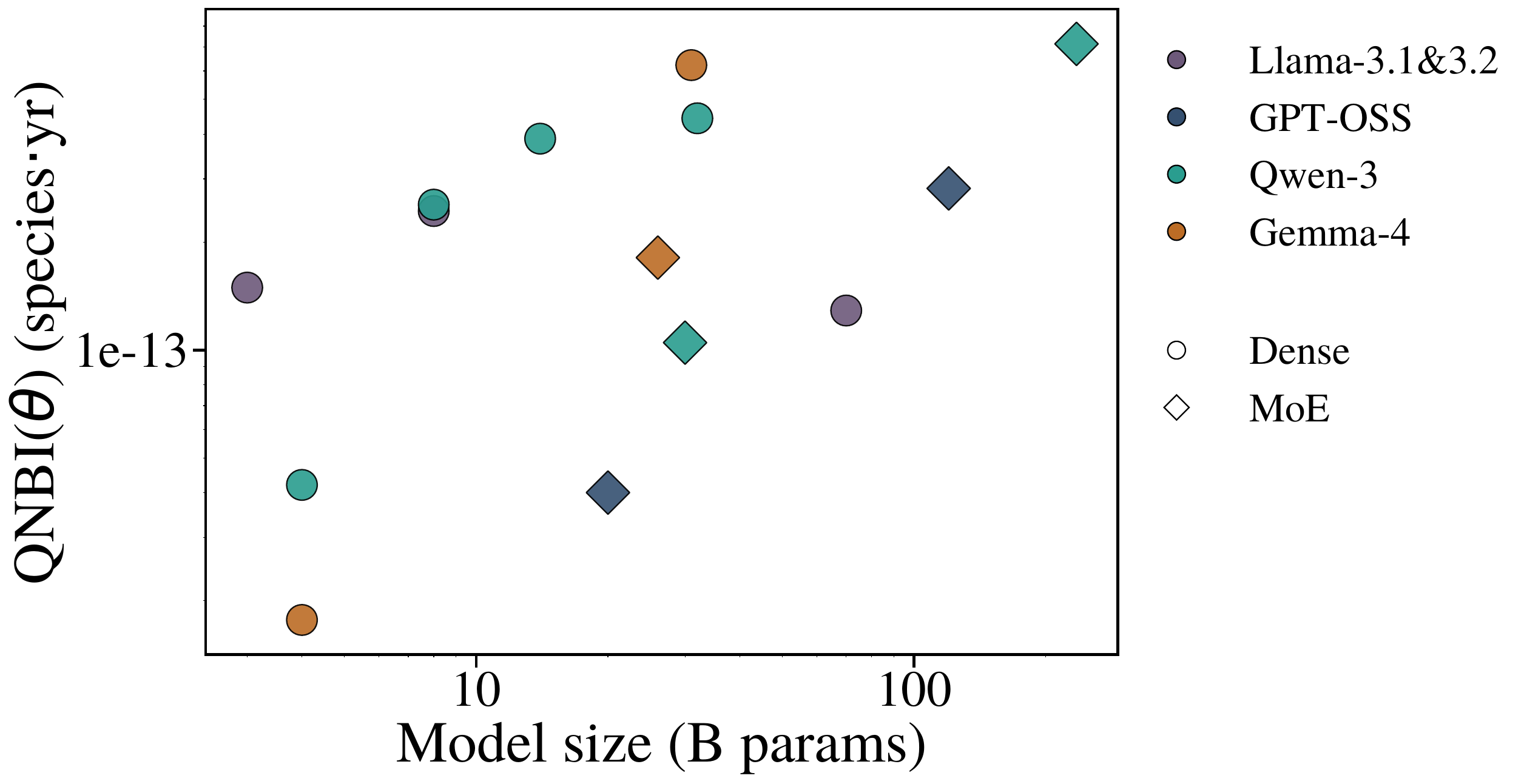}
    \caption{QNBI for MMLU-Pro.}
    \label{fig:app_mmlu_pro_qnbi}
\end{figure}

\begin{figure}[t]
    \centering
    \includegraphics[width=\linewidth]{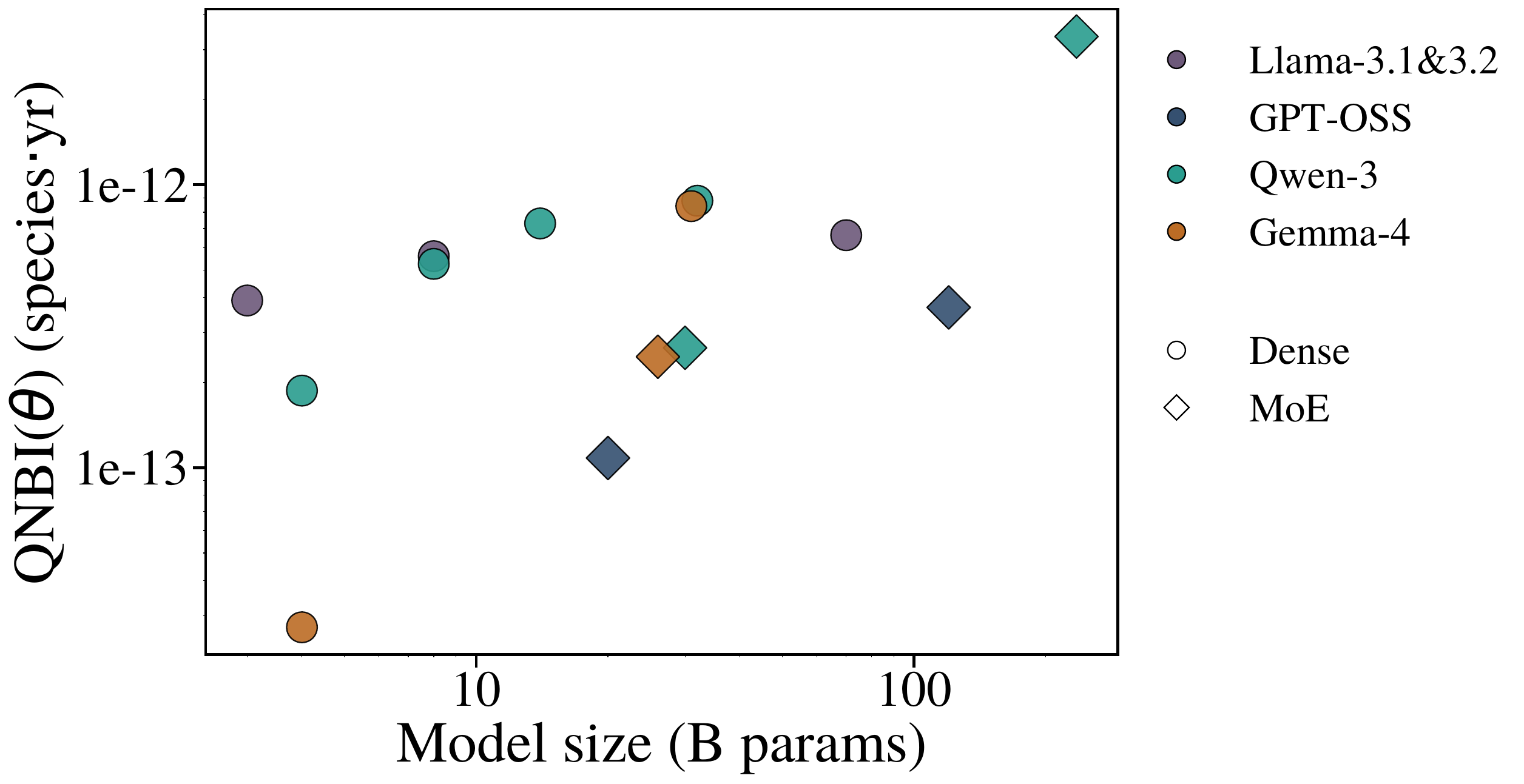}
    \caption{QNBI for SuperGPQA.}
    \label{fig:app_supergpqa_qnbi}
\end{figure}

\begin{figure}[t]
    \centering
    \includegraphics[width=\linewidth]{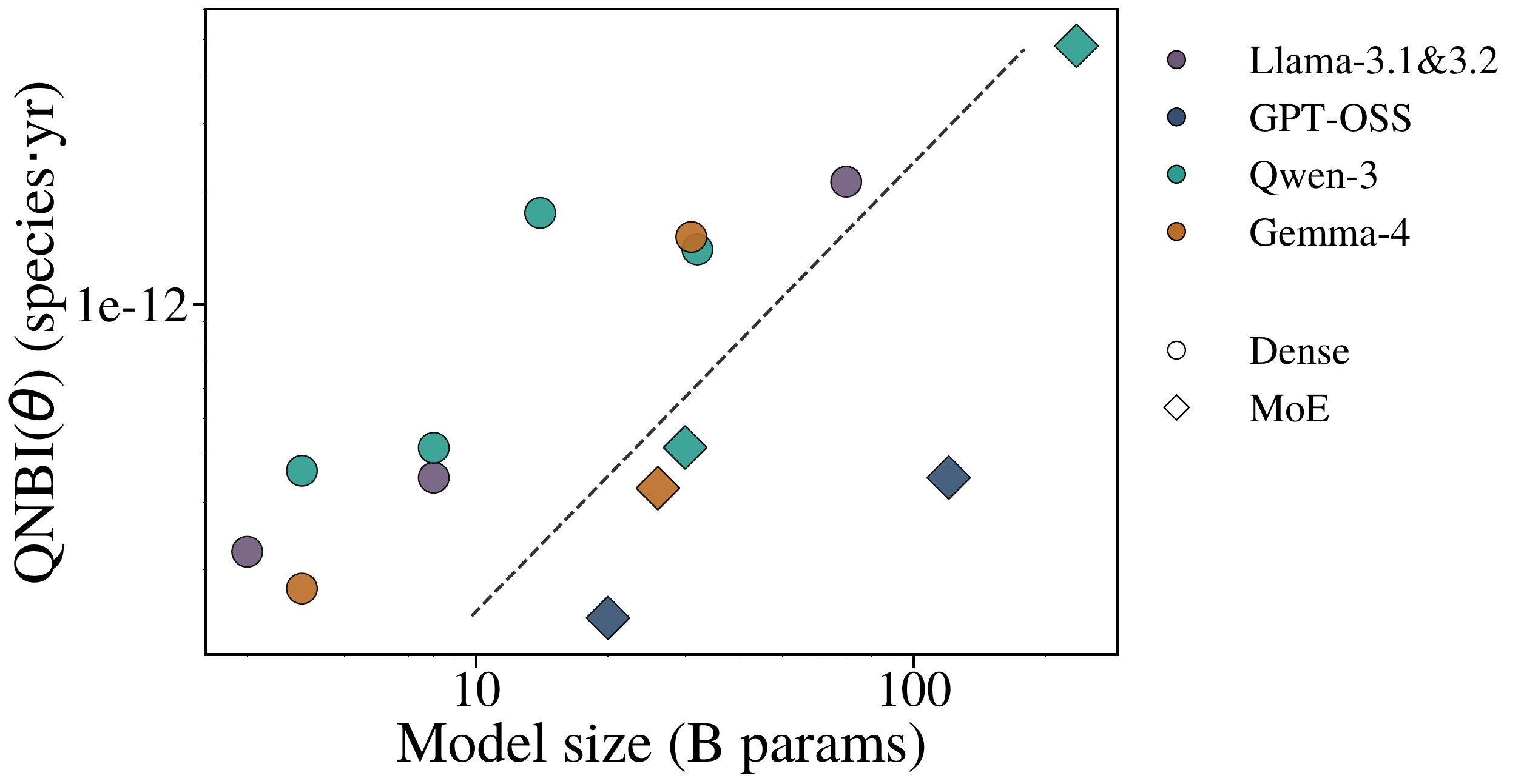}
    \caption{QNBI for LongBench long-output summarization.}
    \label{fig:app_lb_long_sum_qnbi}
\end{figure}

\begin{figure}[t]
    \centering
    \includegraphics[width=\linewidth]{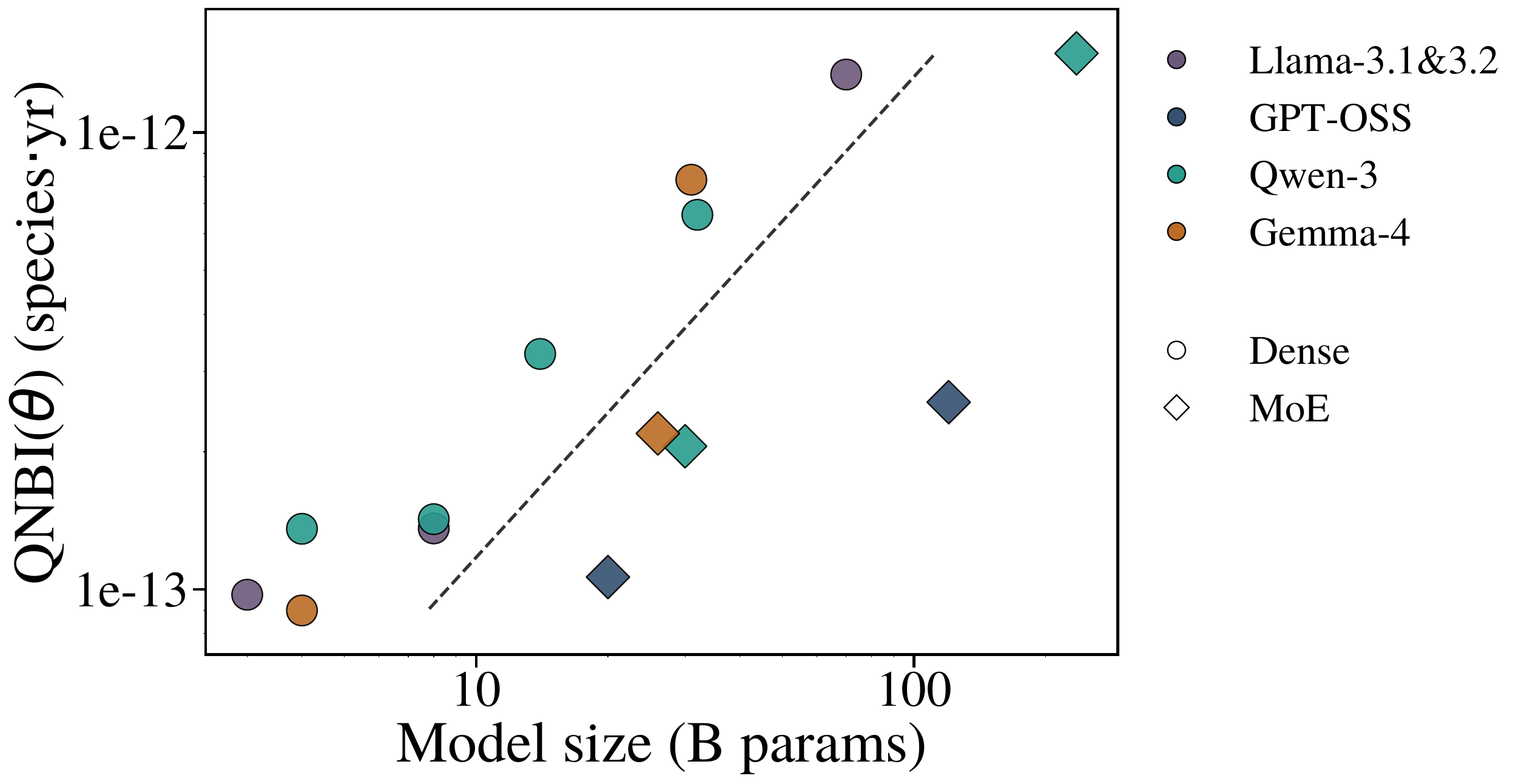}
    \caption{QNBI for LongBench medium-output summarization.}
    \label{fig:app_lb_medium_sum_qnbi}
\end{figure}

\begin{figure}[t]
    \centering
    \includegraphics[width=\linewidth]{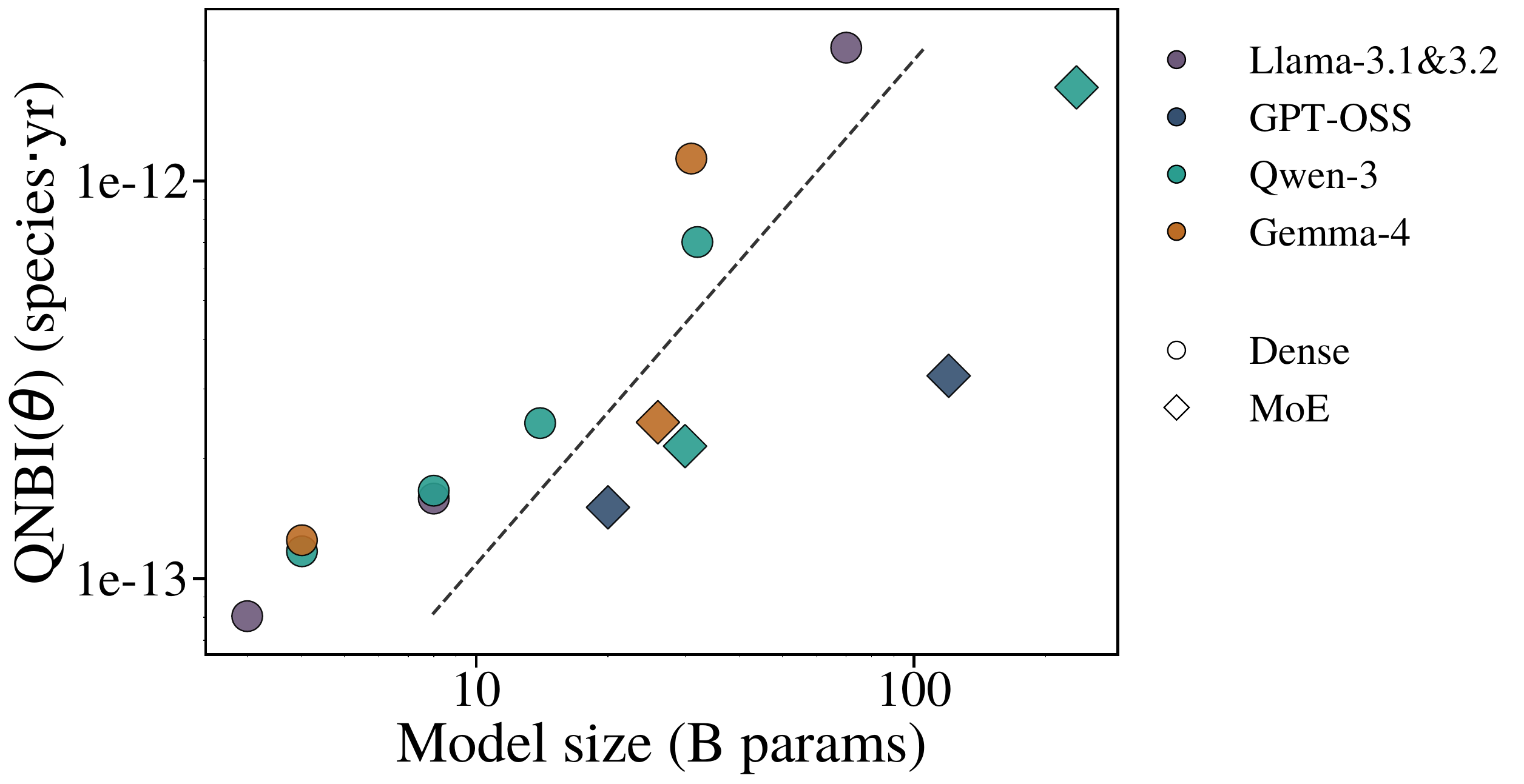}
    \caption{QNBI for LongBench medium-answer RAG QA.}
    \label{fig:app_lb_rag_qa_qnbi}
\end{figure}

\begin{figure}[t]
    \centering
    \includegraphics[width=\linewidth]{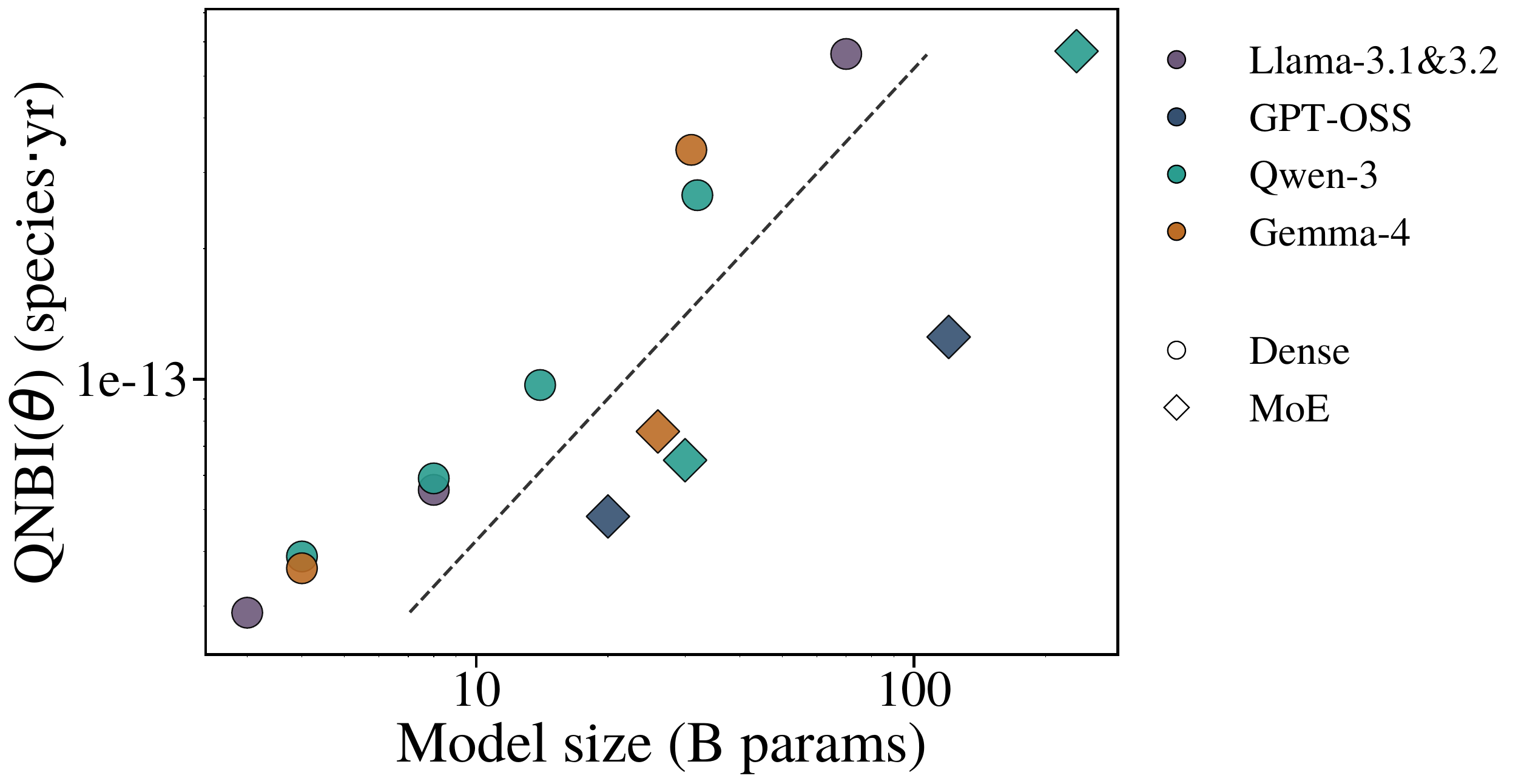}
    \caption{QNBI for LongBench short-answer document QA.}
    \label{fig:app_lb_short_docqa_qnbi}
\end{figure}

This section provides additional results following the same visualization style of \Cref{fig:raw_bi_qscore}, \Cref{fig:qnbi_scatter_sharegpt}, and \Cref{fig:bi_gpu}.
For code-completion workloads, \Cref{fig:app_crosscodeeval_bi,fig:app_crosscodeeval_qnbi} report CrossCodeEval results and \Cref{fig:app_repobench_bi,fig:app_repobench_qnbi} report RepoBench results.
For reasoning workloads, \Cref{fig:app_mmlu_pro_bi,fig:app_mmlu_pro_qnbi} report MMLU-Pro results and \Cref{fig:app_supergpqa_bi,fig:app_supergpqa_qnbi} report SuperGPQA results.
For LongBench workloads, \Cref{fig:app_lb_long_sum_bi,fig:app_lb_long_sum_qnbi,fig:app_lb_medium_sum_bi,fig:app_lb_medium_sum_qnbi,fig:app_lb_rag_qa_bi,fig:app_lb_rag_qa_qnbi,fig:app_lb_short_docqa_bi,fig:app_lb_short_docqa_qnbi} report the four evaluated task buckets.

Across these additional workloads, the overall BI--quality trend is consistent with the main-text ShareGPT case: larger models generally improve quality while increasing BI$_\text{fu}$, and QNBI often favors intermediate-scale or sparse models rather than the largest configuration. The $\sim$30B MoE models frequently form a BI$_\text{fu}$ valley, reflecting their ability to provide higher quality without activating all parameters per token. Gemma4-26B-A4B, however, often incurs higher BI$_\text{fu}$ than Gemma4-E4B, especially on longer-context tasks where memory pressure and longer serving time increase per-request energy. In the size--QNBI space, MoE and dense models remain mostly separable, with MoE models usually achieving lower QNBI at comparable total model size. RepoBench and reasoning workloads are weaker exceptions, where longer prompts or responses may reduce the sparse-activation advantage by increasing serving overhead and expert activity. LongBench summarization quality scores do not show a clear trend due to the reason mentioned in \Cref{app:output_quality}. 

\begin{figure}[t]
    \centering
    \includegraphics[width=\linewidth]{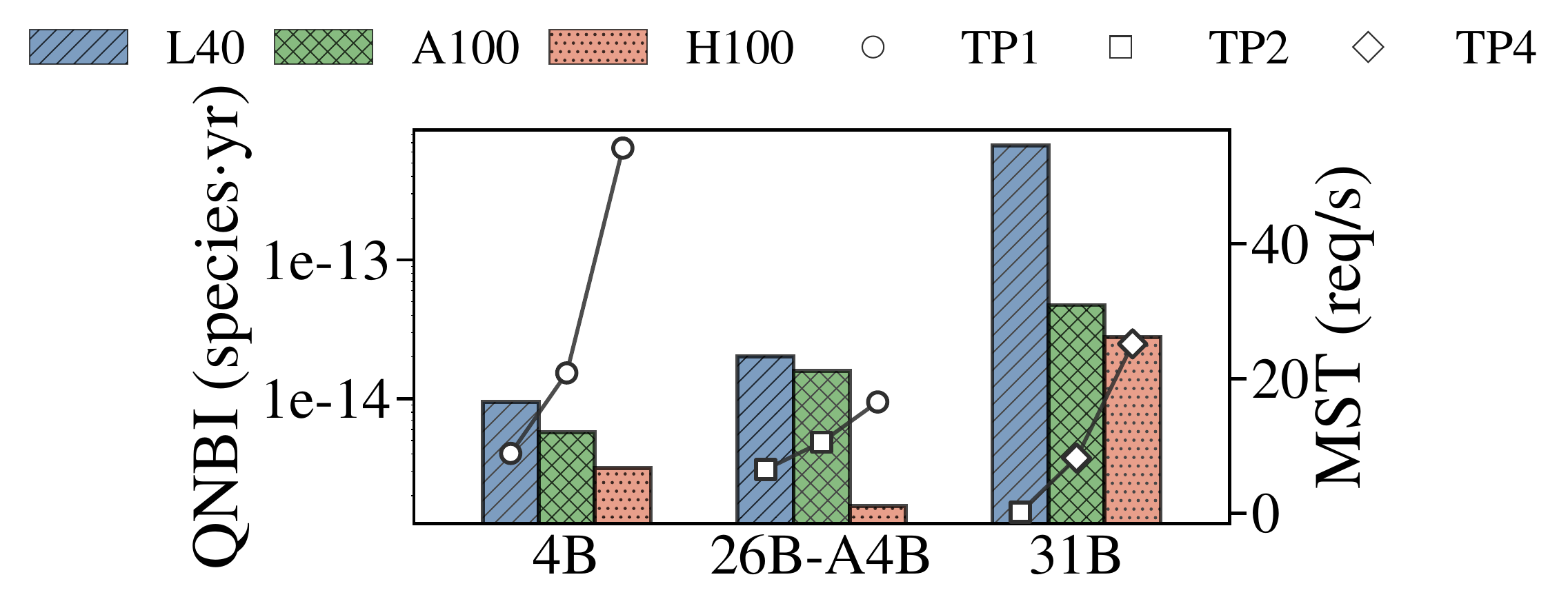}
    \caption{GPU-generation effect on QNBI for Gemma models on ShareGPT.}
    \label{fig:app_gemma_gpu_qnbi}
\end{figure}

Finally, the Gemma GPU-generation study in \Cref{fig:app_gemma_gpu_qnbi} follows the same pattern as the Qwen3 case: newer GPUs reduce QNBI by sustaining higher throughput and avoiding inefficient multi-GPU execution.

\end{appendix}